\documentclass[aps,preprint,groupedaddress,showpacs]{revtex4}
\usepackage{graphicx}
\begin{document}
\bibliographystyle{apsrev}


\title{Evidence for the double degeneracy of the ground-state in the 3D
$\pm J$ spin glass}



\author{Naomichi Hatano}
\email[]{hatano@phys.aoyama.ac.jp}
\homepage[]{http://www.phys.aoyama.ac.jp/~hatano}
\affiliation{Center for Nonlinear Science and Theoretical Division\\
  Los Alamos National Laboratory, Los Alamos, NM 87545}
\affiliation{
Department of Physics, Aoyama Gakuin University,
Setagaya, Tokyo 157-8572, Japan}
\author{J. E. Gubernatis}
\affiliation{Center for Nonlinear Science and Theoretical Division\\
  Los Alamos National Laboratory, Los Alamos, NM 87545}


\date{\today}

\begin{abstract}
A bivariate version of the multicanonical Monte Carlo method and its
application to the simulation of the three-dimensional $\pm J$ Ising
spin glass are described.  We found the autocorrelation time
associated with this particular multicanonical method was
approximately proportional to the system volume, which is a great
improvement over previous methods applied to spin-glass simulations.
The principal advantage of this version of the multicanonical method,
however, was its ability to access information predictive of
low-temperature behavior. At low temperatures we found results on the
three-dimensional $\pm J$ Ising spin glass consistent with a double
degeneracy of the ground-state: the
order-parameter distribution function $P(q)$ converged to two
delta-function peaks and the Binder parameter approached unity as the
system size was increased.  With the same density of states used to
compute these properties at low temperature, we found their behavior
changing as the temperature is increased towards the spin glass
transition temperature. Just below this temperature, the behavior is
consistent with the standard mean-field picture that has an infinitely
degenerate ground state. Using the concept of zero-energy
droplets, we also discuss the structure of the ground-state
degeneracy.  The size distribution of the zero-energy droplets was
found to produce the two delta-function peaks of $P(q)$.
\end{abstract}

\maketitle

\section{Introduction}\label{sec1}
Although a quarter century has passed since the pioneering work on
spin glasses by Edwards and Anderson~\cite{EA75}, many fundamental
problems remain unsolved even for the simplest
models of these materials~\cite{Binder-review}.  In the present
paper, we report the results of Monte Carlo simulations of
the three-dimensional $\pm J$ Ising spin glass, focusing on the nature
of the low-temperature phase~\cite{Hatano-SG1,Gubernatis00}. We
believe we have found important evidence of a doubly-degenerate
ground-state. With the same density of states used to
compute this evidence, we found this character
changing as the temperature is increased towards the spin glass
transition temperature. Just below this temperature, the behavior is
consistent with the standard mean-field picture that has an infinitely
degenerate ground state.

The spin-glass state is characterized by randomly quenched exchange
interactions with frustration. Often systems exhibiting this state
are modeled by the $\pm J$ Ising model. In three dimensions this model
is defined by the Hamiltonian~\cite{EA75}
\begin{equation}\label{eq2-10}
{\cal H}\equiv
-\sum_{\left\langle i,j \right\rangle}
J_{ij}\sigma_{i}\sigma_{j},
\end{equation}
where the Ising spins $\sigma_{i}$ fluctuate thermodynamically, while
each exchange interaction $J_{ij}$ is quenched to $\pm J$ randomly.
The summation runs over the nearest-neighbor pair of sites on a cubic
lattice.

The frustration of local spin configurations by the exchange
interactions generates many local minima in the free energy landscape,
each minimum representing a seemingly random configuration of the
spins.  At low temperatures, the spins, however, may freeze into
some configuration, and this freezing is the essence of the
spin-glass ``order.''  Because of the nature of this
order, equilibration of spin glasses in simulations and experiments
is often very hard.

\subsection{Nature of the spin-glass phase}

It is becoming increasingly accepted that the $\pm J$ model has a
finite-temperature spin-glass phase transition in three dimensions,
particularly after several recent Monte Carlo
studies~\cite{Kawashima,Hukushima,pal_and_car,Ballesteros}. The nature 
of the
low-temperature
phase, however, is still controversial. Historically, the controversy
has been mainly
between advocates of the
mean-field picture and the droplet picture.  The mean-field advocates
maintain
the existence of an infinite number of global minima of the free
energy in the low-temperature phase. The condition is called replica
symmetry breaking~\cite{RSB}. The breaking of this symmetry is
rigorously true for the mean-field (Sherrington-Kirkpatrick)
model~\cite{SK75,Parisi79} of spin glasses. The question is
whether it is also found in finite-dimensional non-mean-field models.
The droplet
advocates~\cite{bray,droplet-letter,droplet} 
on the other hand,
assert that the free-energy
landscape has only two global minima that are connected through spin
inversion symmetry.  In the
droplet picture, the nature of the three-dimensional ground state is
seemingly less exotic than in the mean-field picture.

The difference in the ground state degeneracy between the two pictures
becomes quantitative when we
use the overlap order parameter of the spin-glass phase~\cite{EA75}.
To define this order parameter, we
replicate the random exchange interactions in (\ref{eq2-10}) and change
the Hamiltonian to
\begin{equation}\label{eq2-20}
{\cal H}\equiv
-\sum_{\alpha=1,2}\sum_{\left\langle i,j \right\rangle}
J_{ij}\sigma_{i}^{(\alpha)}\sigma_{j}^{(\alpha)},
\end{equation}
where the superscript on the spin variables labels a replica.  The
overlap order parameter is then defined as the
Hamming distance between spin configurations of the
two replicas~\cite{EA75}:
\begin{equation}\label{eq2-30}
q=\frac{1}{L^d}\sum_{i=1}^{L^d}\sigma_{i}^{(1)}\sigma_{i}^{(2)}.
\end{equation}
Here $L$ is the linear size of the system and $d$ is the
dimensionality, which is three throughout the paper.

The spins of one replica and those of the other are thermodynamically
independent.  Still they can be correlated because of the common
random exchange interactions.  In the high-temperature limit, the spin
configurations are uncorrelated, and hence the overlap order parameter
tends to zero as $L^{-d/2}$.  In the low-temperature limit, on the
other hand, the spin configurations are frozen into energy minima, and
hence the overlap order parameter can have a finite value.

In the droplet picture, the overlap order parameter at low
temperatures and in the thermodynamic limit can take only two values
equal and opposite to each other, as the frozen spin configuration of
one replica is either macroscopically identical to the configuration
of the other replica or to its inverted configuration.  (We note,
however, that the identity needs only to be macroscopic; the
configurations can differ locally.)  Hence the overlap order parameter $q$
takes one of the two values with an equal probability.  In the
mean-field picture, on the other hand, the overlap order parameter can
take various values.  Because of the many free-energy minima, the
frozen spin configurations of the two replicas can be macroscopically 
different.

More explicitly, we define the order-parameter distribution as
\begin{equation}\label{eq-pofq}
P(q)\equiv
\left[  \frac{1}{\cal Z}
\int  D(E,q) e^{-\beta E }dE
\right]_{\mathrm{av.}},
\end{equation}
where the partition function is given by
\begin{equation}\label{eq-partfn}
{\cal Z}\equiv
\sum_{\{\sigma\}}e^{-\beta E(\{\sigma\})}
=\int dq\int dE D(E,q) e^{-\beta E}.
\end{equation}
In these two equations $D(E,q)$ is the normalized density of states
for the energy $E$ and the overlap order parameter $q$.
The energy in Eqs.~(\ref{eq-pofq}) and (\ref{eq-partfn}) and other
energies hereafter
(including those in the Monte Carlo simulations) are the energies of the
replica
Hamiltonians (\ref{eq2-20}).  The square brackets in
Eq.~(\ref{eq-pofq}) denote the random average over various samples
$\{J_{ij}\}$.
The distribution $P(q)$ is normalized so that
\begin{equation}
\int P(q)dq=\frac{2}{L3}\sum_{q=-1}^{q=1}P(q)=1.
\end{equation}
Physical quantities that are functions of $q$ only are
calculated as
\begin{equation}
\left[
\left\langle
f(q)
\right\rangle
\right]_{\mathrm{av.}}
=\int f(q)P(q) dq,
\end{equation}
where the angular brackets denote the thermodynamic average;
{\it e.g.\/}, the spin-glass susceptibility
\begin{equation}\label{eq-chisg}
\chi_{\mathrm{sg}}=
L^d\beta2 \left[
\left\langle q2 \right\rangle
- \left\langle q \right\rangle2
\right]_{\mathrm{av.}}
\end{equation}
and the Binder parameter
\begin{equation}\label{eq-binder}
g_{\mathrm{sg}}(T,L)\equiv\frac{3}{2}\left(
1-
\frac{\displaystyle \left[\left\langle
q4\right\rangle\right]_{\mathrm{av.}}}%
{\displaystyle 3\left[\left\langle q2\right\rangle\right]_{\mathrm{av.}}^2}
\right).
\end{equation}

\begin{figure}
\begin{center}
\includegraphics[width=8cm]{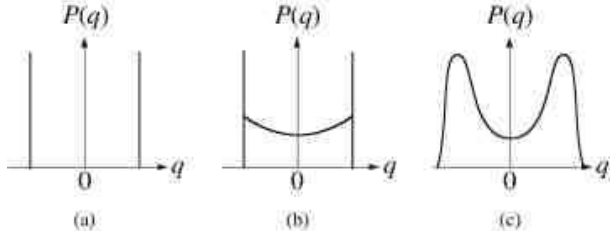}
\end{center}
\caption{The functional form of the order-parameter distribution
$P(q)$ in the low-temperature phase:
(a) According to the droplet picture;
(b) According to the mean-field picture;
(c) For finite-sized systems.}
\label{fig-pofq-schem}
\end{figure}

The droplet picture and the mean-field pictures have the
functional forms of $P(q)$ as shown in Fig.~\ref{fig-pofq-schem}.
In the droplet picture, the order-parameter distribution  in the
thermodynamic limit has two
delta-function peaks, indicating two
possible thermodynamic states (Fig.~\ref{fig-pofq-schem}a).
In the mean-field picture the
order-parameter distribution is continuous between these peaks
(Fig.~\ref{fig-pofq-schem}b).

In practice, we can only simulate finite systems for which
the order-parameter distribution looks like Fig.~\ref{fig-pofq-schem}c.
We need to see whether the distribution in Fig.~\ref{fig-pofq-schem}c
converges to Fig.~\ref{fig-pofq-schem}a or Fig.~\ref{fig-pofq-schem}b as
$L\to\infty$. If $\lim_{L\to\infty} P(0)\rightarrow 0$, then
significant doubt is cast
upon the mean-field picture.

While some simulations appear to support infinite
degeneracy~\cite{Marinari96,Inigues97,Marinari99},
others do not
\cite{Hartmann00,Palassini99,Palassini00,Palassini01,komori}, and a
recent analytic
treatment argues against it~\cite{Newman96,Newman98}.  We note that
the droplet picture is a theory concerning the {\em zero}-temperature
fixed point, while most studies suggesting the validity of the
mean-field picture are based on numerical estimates of $P(0)$ at
$T$ only as low as $0.7T_c$, where the glass transition temperature $T_c$
is approximately 1.0. If the doubly degenerate ground state predicted
by droplet picture is correct, it should at
least be seen at temperatures lower than 0.7$T_c$, the closer to $T=0$
the better.  Unfortunately, the slow dynamics of spin glasses has
hindered numerical simulations from exploring the vicinity of the
zero-temperature fixed point.  We comment that a recent numerical
study~\cite{Moore98} demonstrated that results of the Migdal-Kadanoff
approximation appear to support the mean-field picture near and below
the glass transition temperature, but eventually support the droplet
picture as $T\to0$.  Uncertainty, however, remains because the
Migdal-Kadanoff approximation favors the droplet picture, even for the
mean-field model \cite{comment1,comment2}.

During the course of our work and just after, the possibility that the
behavior of the spin glass does not unequivocally fit into one of these
two pictures over the entire temperature range began to be
appreciated. For example, Krzakala and Martin~\cite{Krzakala01}
proposed the novel perspective that the entropy fluctuations lead to a
trivial $P(0)$ at zero temperatures, even if there are zero-energy
large-scale excitations (complex energy landscape). They further
proposed that such a situation should arise in the the
three-dimensional $\pm J$ Ising spin glass, and argued that if the
energy landscape is complex with a finite number of ground state
families, then replica symmetry breaking reappears at finite
temperatures. This perspective contested the standard picture.
Palassini and Young~\cite{Palassini99,Palassini00,Palassini01}
studied this scenario and concluded that the size dependence of $P(q)$
around $q=0$ is trivial and does not support the ultrametric picture.

Katzgraber, Palassini, and Young~\cite{Katzgraber01a,Katzgraber01b}
addressed the
issue of Gaussian versus bimodal distribution by simulating at finite
temperatures the
Gaussian exchange model and concluded that as in the bimodal exchange
model $P(0)$ is trivial in the thermodynamic limit and suggested the
existence of low finite energy excitations that cost finite energy and whose
surface has fractal dimension less than the spatial dimension of the
system. However, the sizes they studied were quite small and so they
concluded there might be a crossover at larger sizes to a different
behavior, such as a droplet or replica symmetry breaking picture.

Two related papers by Hed et al. \cite{domany1,domany2}
suggest that the spin-glass phase possesses some characteristics of
the mean-field description as a non-trivial $P(0)$ and a hierarchical
(but not ultrametric) structure of the pure states; nevertheless, they
also claim that this phase is consistent with the Fisher-Huse
scenario of the droplet picture. Correlated spin domains serve as the
cores of zero enegy excitations.

The recent work of Marinari {\it et al}.~\cite{Marinari01} adopted the
cluster analysis of Hed {\it et al}.~\cite{Hed01} and found strong
continuity among physical features for $T>0$ and those found at $T=0$,
leading to a scenario with emerging mean-field like characteristics
that are enhanced in the large volume limit for $T>0$. These
mean-field like features arise with entropic fluctuations. More
recently Lamarcq {\it et al}.~\cite{Lamarcq01} have studied the
fractal dimension of the clusters that are the low-lying excitations
of the model.

There are still other papers, for example,
\cite{Houdayer1,Houdayer2,stein,nishimori}, all illustrating a shift
from simply the droplet versus the mean-field picture to something
more subtle, with a consensus still very much evolving. One focus is
on the nature and geometrical structure of ground-state excitations.

\subsection{Present Study}

In the present study, we significantly reduced the difficulty of the
slow dynamics in Monte Carlo simulations by using a bivariate version
of Berg and Neuhaus's multicanonical Monte Carlo method.
Multicanonical simulations are performed independent of temperature or
of a range of nearby temperatures and estimate the density of
states. From it we can in principle calculate expectation values at
any temperatures.  Our resulting estimates of $P(0)$ at low
temperatures and as a function of lattice size suggest that at very
low temperatures thermodynamic behavior could be consistent with a
doubly degenerate ground state, while at higher temperatures, it is
consistent with an infinite degenerate ground state. Hence, we found
that an intrinsic temperature-independent quantity seemingly exhibited
different looking equilibrium behavior at different temperatures.

We present our Monte Carlo results in Sect.~\ref{sec2}.  The
order-parameter distribution function $P(q)$ at the low temperature of
$T=0.3 $ exhibits features indicative of a doubly degenerate ground
state: $P(0)$ decreases as the system size is increased.  The
low-temperature behavior of the Binder parameter also suggests double
degeneracy. In Sect.~\ref{sec3} we discuss further implications of
our results for the ground-state degeneracy.
In Appendix, we describe our simulation
method, namely a bivariate multicanonical Monte Carlo method.
Monovariate multicanonical methods~\cite{Berg91,Berg92} have been
applied to spin glasses before~\cite{Berg94,Berg96,Berg98}; we found,
however, that by using a bivariate version we could reduce the
correlation time of the simulation significantly.  We show that the
autocorrelation time is approximately proportional to the system size.

\section{Numerical Results}\label{sec2}

We have carried out a bivariate multicanonical Monte Carlo simulation
of cubic systems with edges $L=4$ (1904 samples), $L=6$ (2843
samples), $L=8$ (1015 samples), and $L=10$ (1111 samples).  The
simulation method directly returns the density of states $D(E,q)$, a
temperature-independent quantity.  We describe the details of the method
and its use in Appendix; however, because the method is subtle, relatively
new,
and quite different from standard methods, we now summarize some of what is
discussed in the Appendix.

With the multicanonical method, sometimes called the entropic sampling
method, we do not equilibrate the simulation at
any value of $T$; that is, we never sample the steady-state
distribution, generated by a Markov chain, that is supposed to
represent the Boltzmann distribution. Instead, we sample from a
steady-state distribution, generated by a Markov chain, that is
adaptively constructed to be flat on the average.  The flatness means
we sample all accessible $(E,q)$ values with an equal probability.
In other words, we sample all {\em thermodynamics} states equally. In
fact, the sampling emphasizes regions where $D(E,q)$ is small and as a
consequence are generally difficult to access with many other
methods.

Sampling from this flat distribution also allows one to
estimate $D(E,q)$. We obtain $D(E,q)$ for different system sizes. Once
we obtain it for a given size, we can in principle obtain
the the properties of the system for any temperature; that is,
properties of the system at different temperature, no matter how
different they may seem, all follow from the same $D(E,q)$.

The validity of our low $T$ predictions depends on the ``flatness''
extending to the ground-state energy. As we discuss more fully in the
Appendix flatness over the entire range of energy is not essential if
not assumed in the evaluation of expectation values. The multicanonical 
method
generally is a good ground-state sampler. This conclusion is supported
in part by our to-be-reported rapid convergence of the entropy of the
ground state for low $T$.  As we will show, our estimates of the
ground state residual entropy are consistent with the ones which
Hartmann obtained by a ground-state counting
algorithm \cite{Hartmann99}.  This agreement is one reason why we are
confident our simulations are covering low-energy states properly.  At
high temperatures, we will also see that the same $D(E,q)$ estimates the
Binder parameter to within two sigmas with the previous results
\cite{Kawashima}.

\subsection{Order-parameter distribution}

\begin{figure}
\begin{center}
\includegraphics[width=8cm]{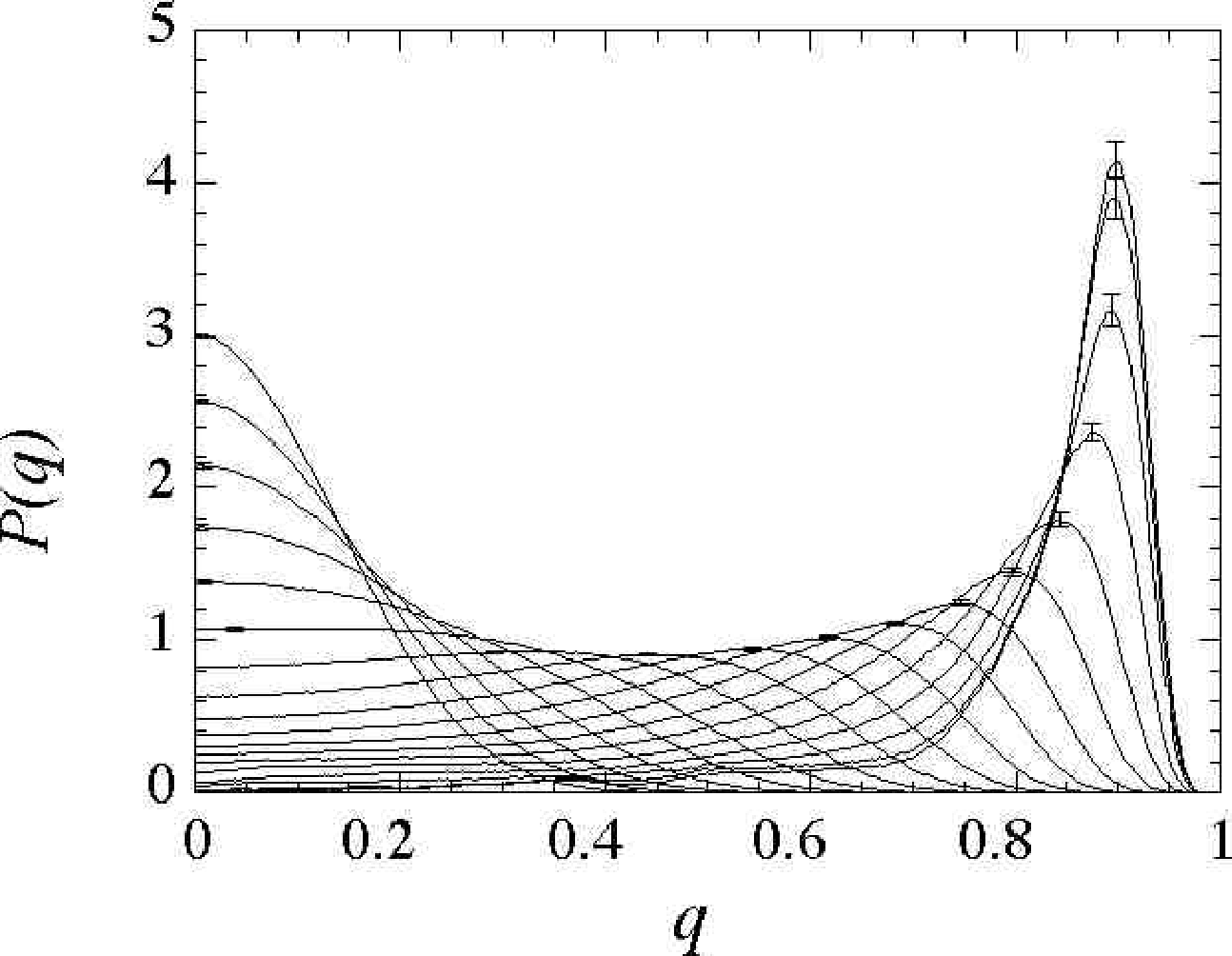}
\end{center}
\caption{The temperature dependence of the order-parameter
distribution $P(q)$ for $L=8$.  The curves with the peak positions
ordered from right to left correspond to $T=0.3, 0.4, 0.5, \ldots,
2.0$.  The statistical errors are indicated only at each peak
position.  We plotted only for $0\leq q
\leq 1$, evoking the symmetry $P(-q)=P(q)$.}
\label{fig-pofqL8}
\end{figure}

\begin{figure}
\begin{minipage}[t]{0.35\textwidth}
\includegraphics[width=\textwidth]{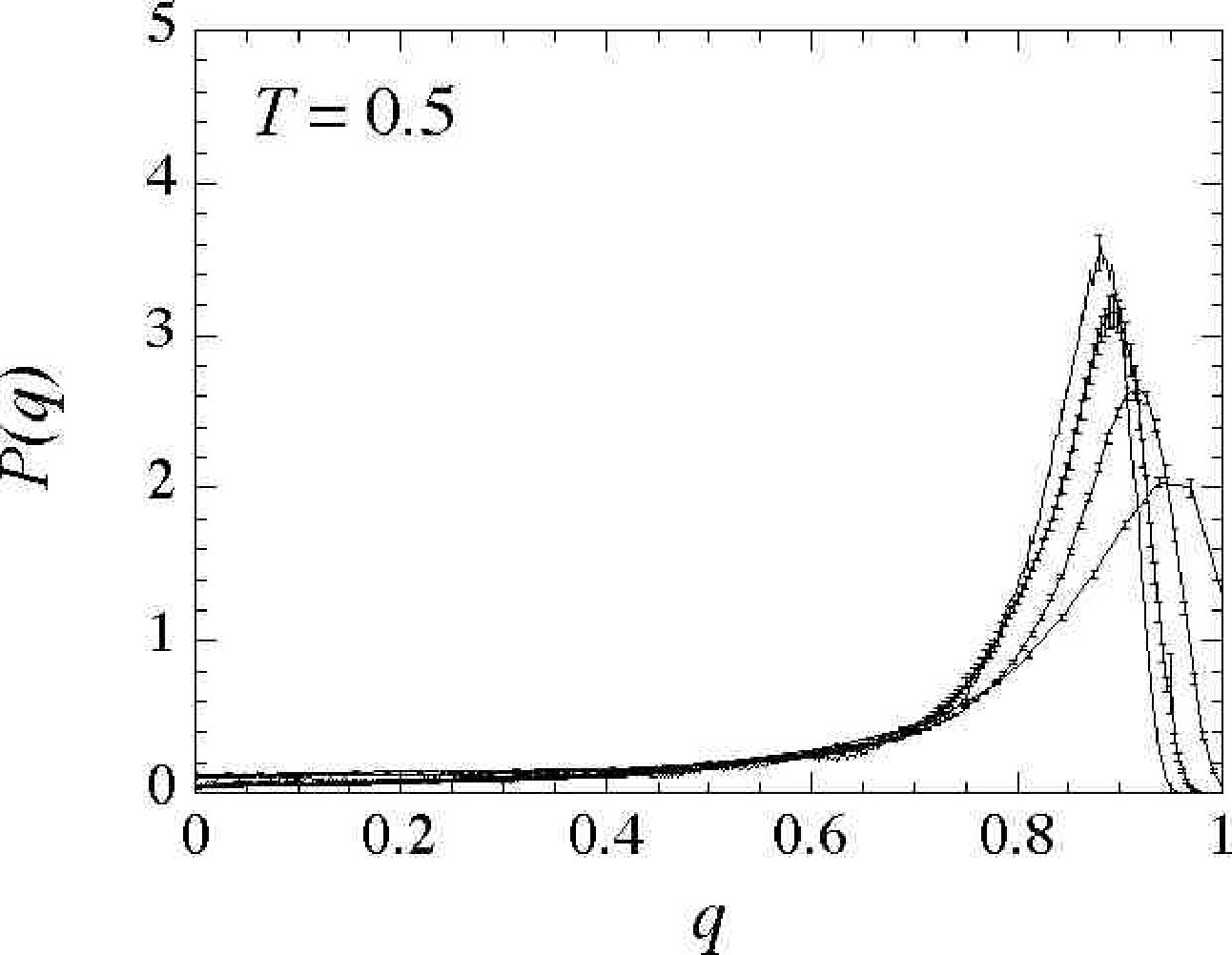}
(a)
\end{minipage}
\hfill
\begin{minipage}[t]{0.35\textwidth}
\includegraphics[width=\textwidth]{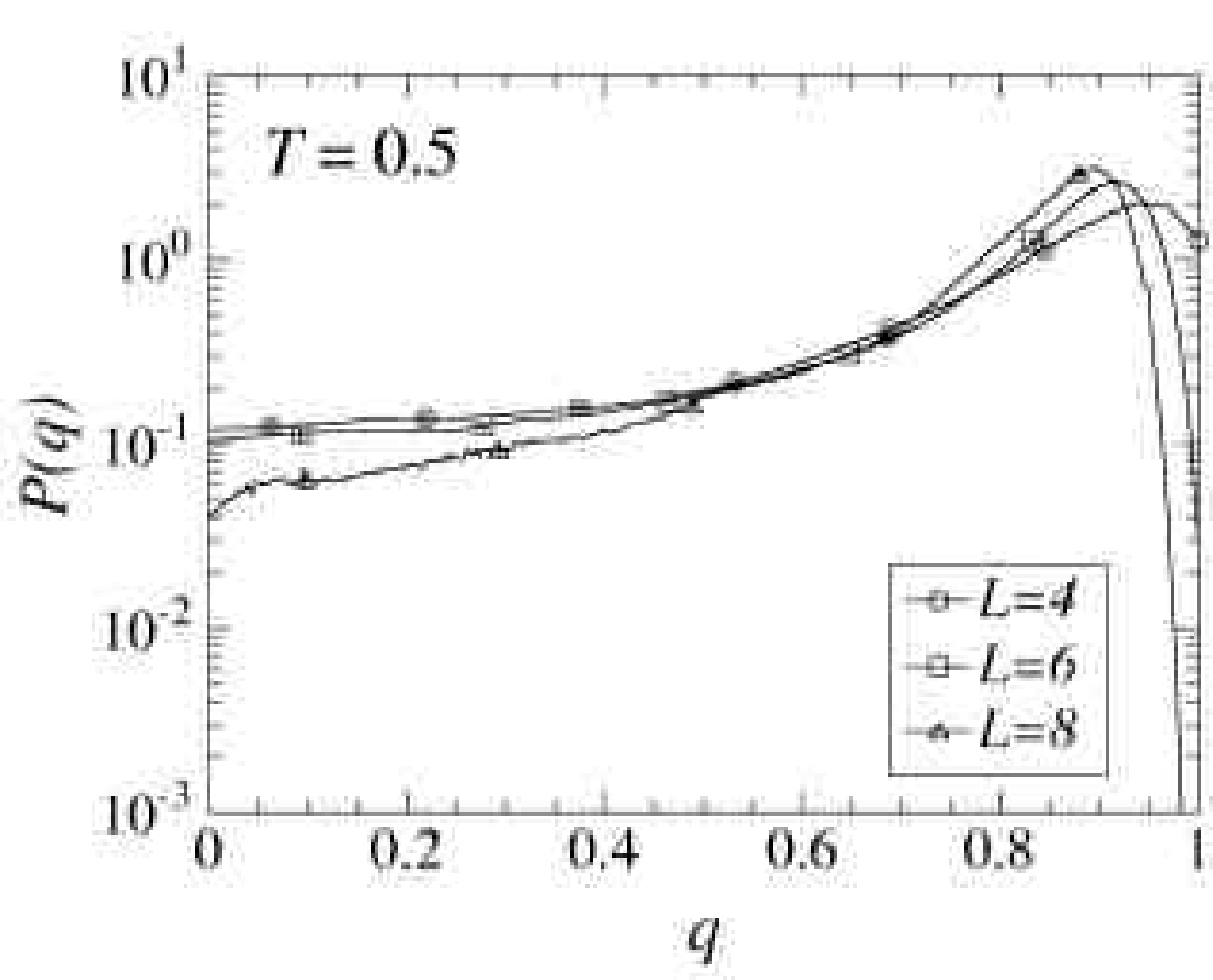}
(b)
\end{minipage}

\bigskip
\begin{minipage}[t]{0.35\textwidth}
\includegraphics[width=\textwidth]{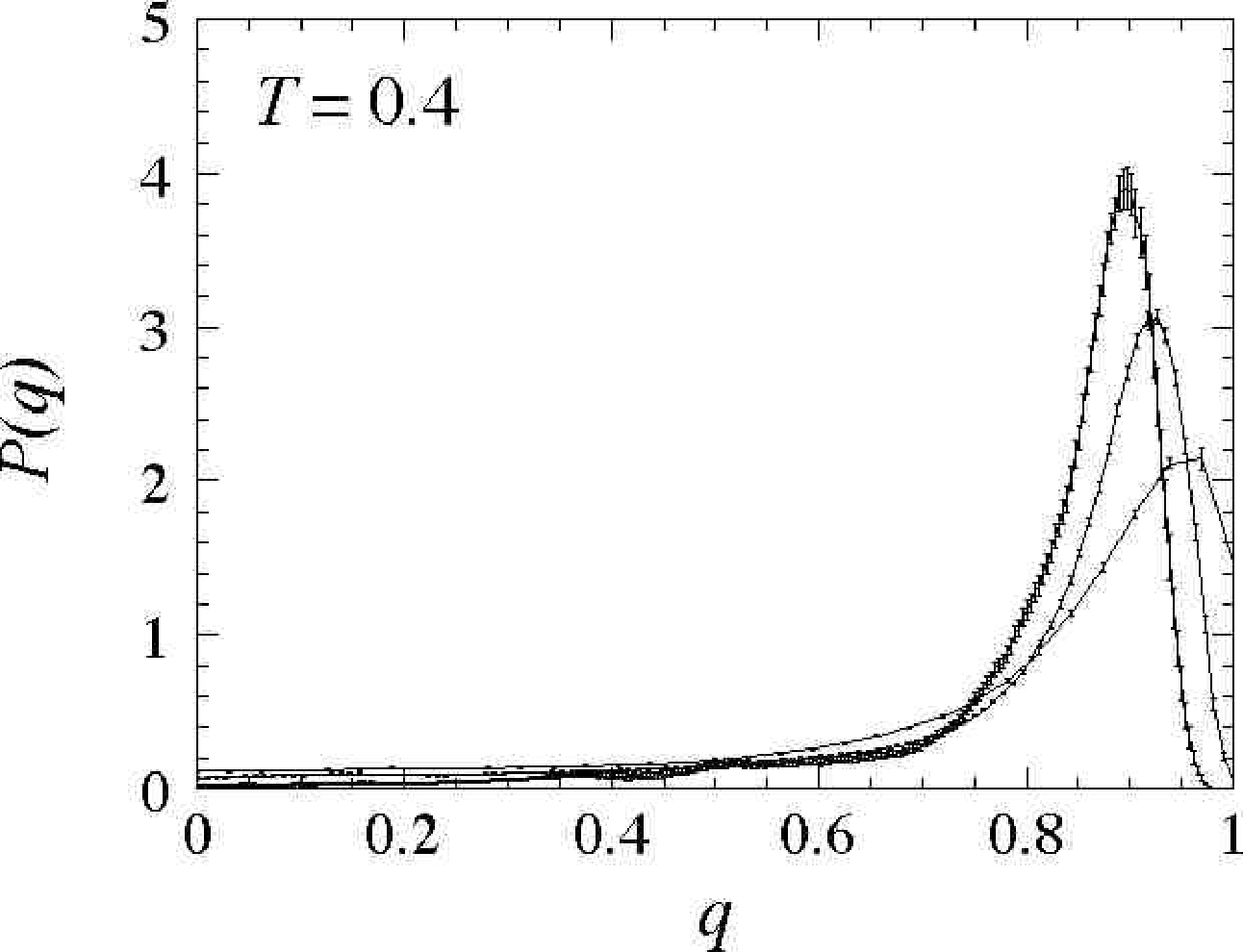}
(c)
\end{minipage}
\hfill
\begin{minipage}[t]{0.35\textwidth}
\includegraphics[width=\textwidth]{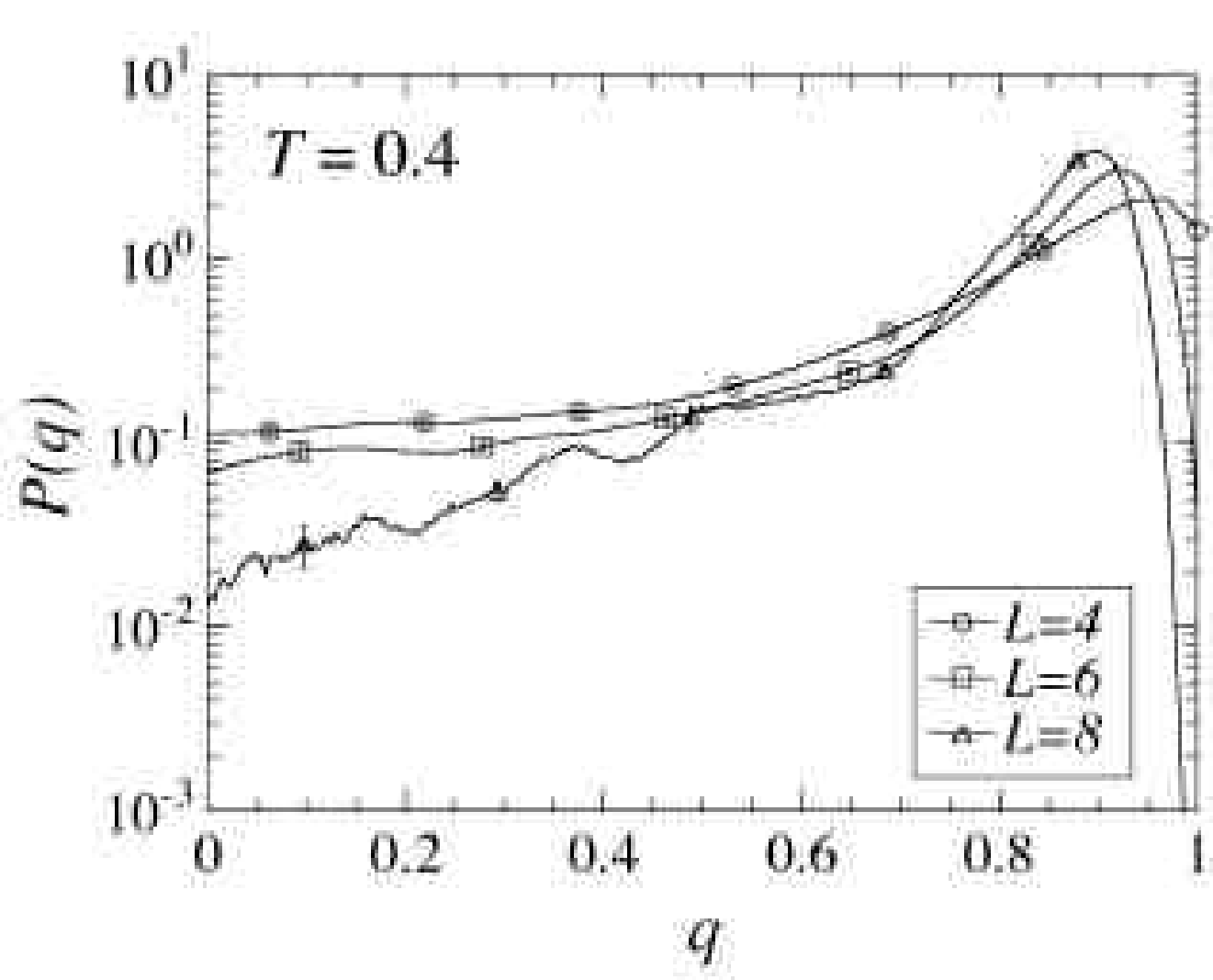}
(d)
\end{minipage}

\bigskip
\begin{minipage}[t]{0.35\textwidth}
\includegraphics[width=\textwidth]{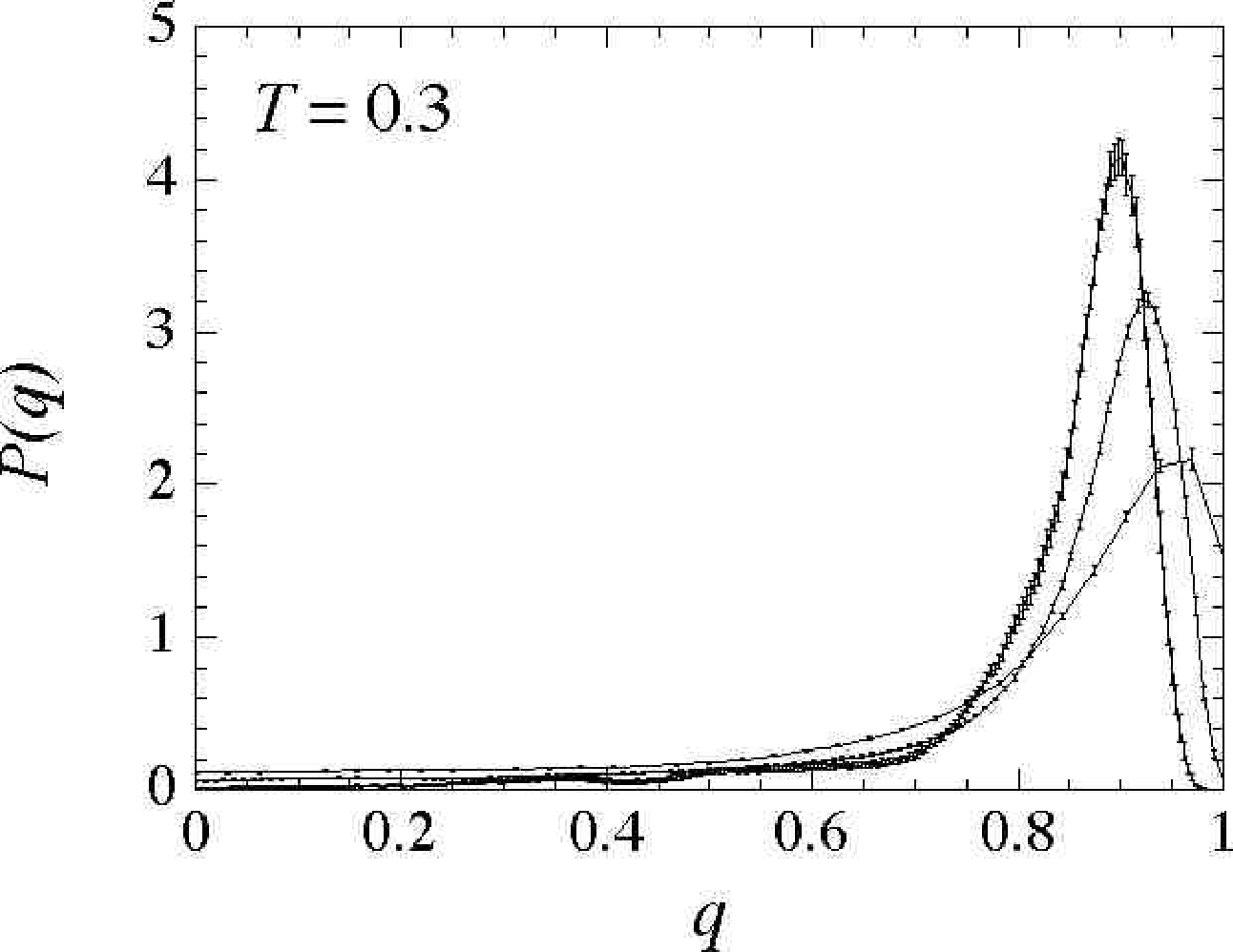}
(e)
\end{minipage}
\hfill
\begin{minipage}[t]{0.35\textwidth}
\includegraphics[width=\textwidth]{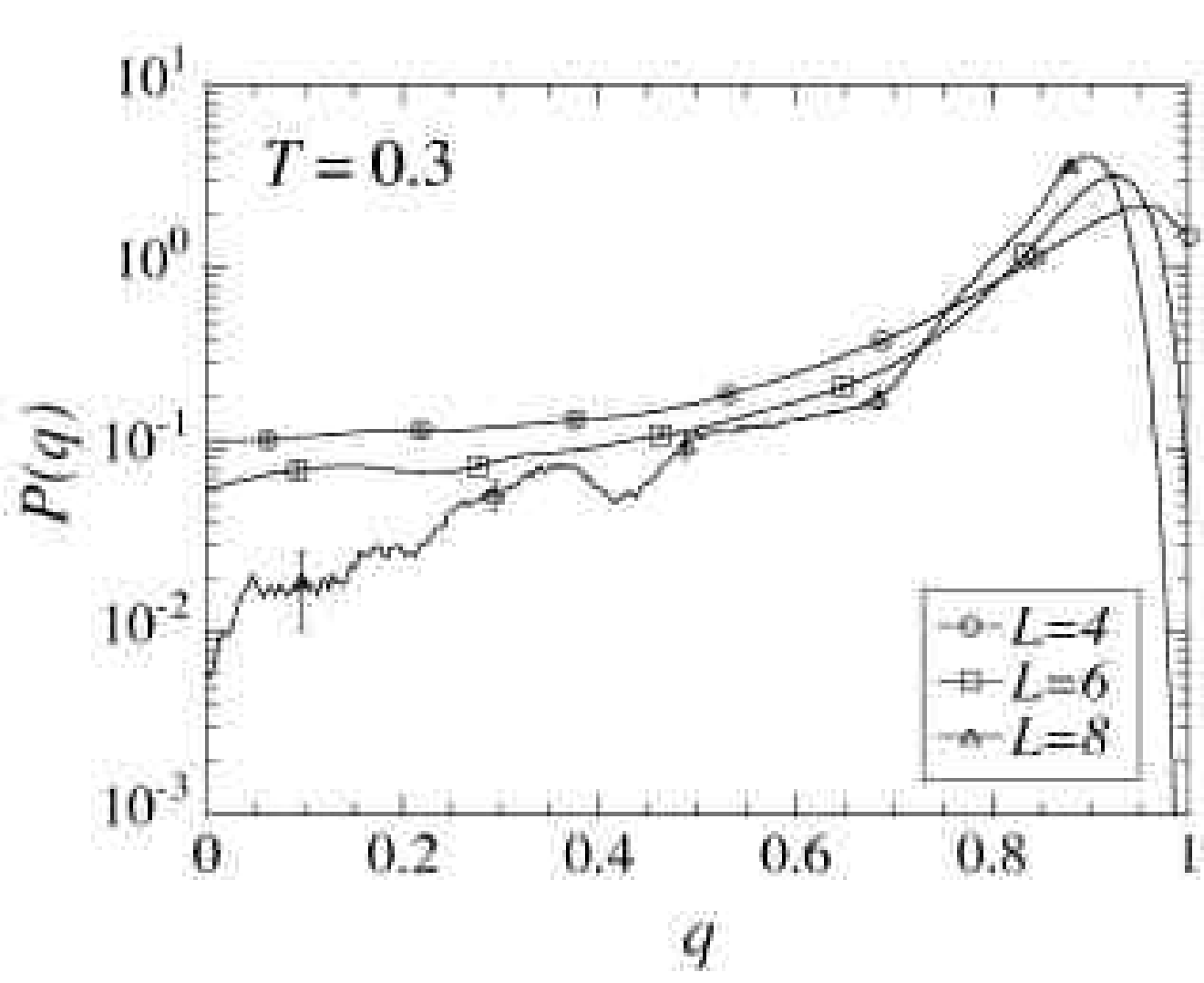}
(f)
\end{minipage}
\caption{
The size dependence of the order-parameter distribution $P(q)$.
(a) A linear plot for $T=0.5$ and for $L=4,6,8$, and $10$.
The peak position moves left as the system size is increased.
Because the data points are very dense for $L=10$, the error bar is
shown only at the peak, where the statistical error is the largest.
(b) A semi-logarithmic plot for $T=0.5$.
The error bars are shown only for a part of the data points.
(c) A linear plot for $T=0.4$ and for $L=4,6$, and $8$.
The peak position moves left as the system size is increased.
(d) A semi-logarithmic plot for $T=0.4$.
The error bars are shown only for a part of the data points.
(e) A linear plot for $T=0.3$ and for $L=4,6$, and $8$.
The peak position moves left as the system size is increased.
(f) A semi-logarithmic plot for $T=0.3$.
The error bars are shown only for a part of the data points.
}
\label{fig-pofq-0.3-0.5}
\end{figure}

   From the density of states $D(E,q)$, we straightforwardly calculated
the order-parameter distribution $P(q)$, following
Eq.~(\ref{eq-pofq}).  Figure~\ref{fig-pofqL8} shows the temperature
dependence of $P(q)$ for $L=8$.  The function is close to a Gaussian
distribution at high temperatures and has a double-peak structure at
low temperatures.
The results for $T=0.3,0.4$, and $0.5$ are shown in
Fig.~\ref{fig-pofq-0.3-0.5} for several values of $L$.
We clearly see the decreasing tendency of $P(0)$ as $L\to\infty$.
The raw data, however, presumably underestimates the true values of
$P(0)$: the $P(q)$ data for $L=8$ noticeably oscillates
for $q\leq0.7$ with the data point at $q=0$ happening to be in a valley of
the oscillation.  We presume that this oscillation is due to
correlations between data at different values of $q$,
for example, the data
point $P(0)$ being correlated with $P(0.01)$; similar
oscillations are seen in the data of others \cite{berg94,Palassini01,Hed01}
that was obtained by quite
different methods.
The present numerical method, as is summarized in Appendix,
generates a random walk in the {\em macroscopic} phase space.
The frequency of access by the random walker may be
statistically less in some area of the phase space.
The density of states will be underestimated in such area.
Thus the correlations between data at different values of $q$ can occur.

\begin{figure}
\begin{center}
\includegraphics[width=0.35\textwidth]{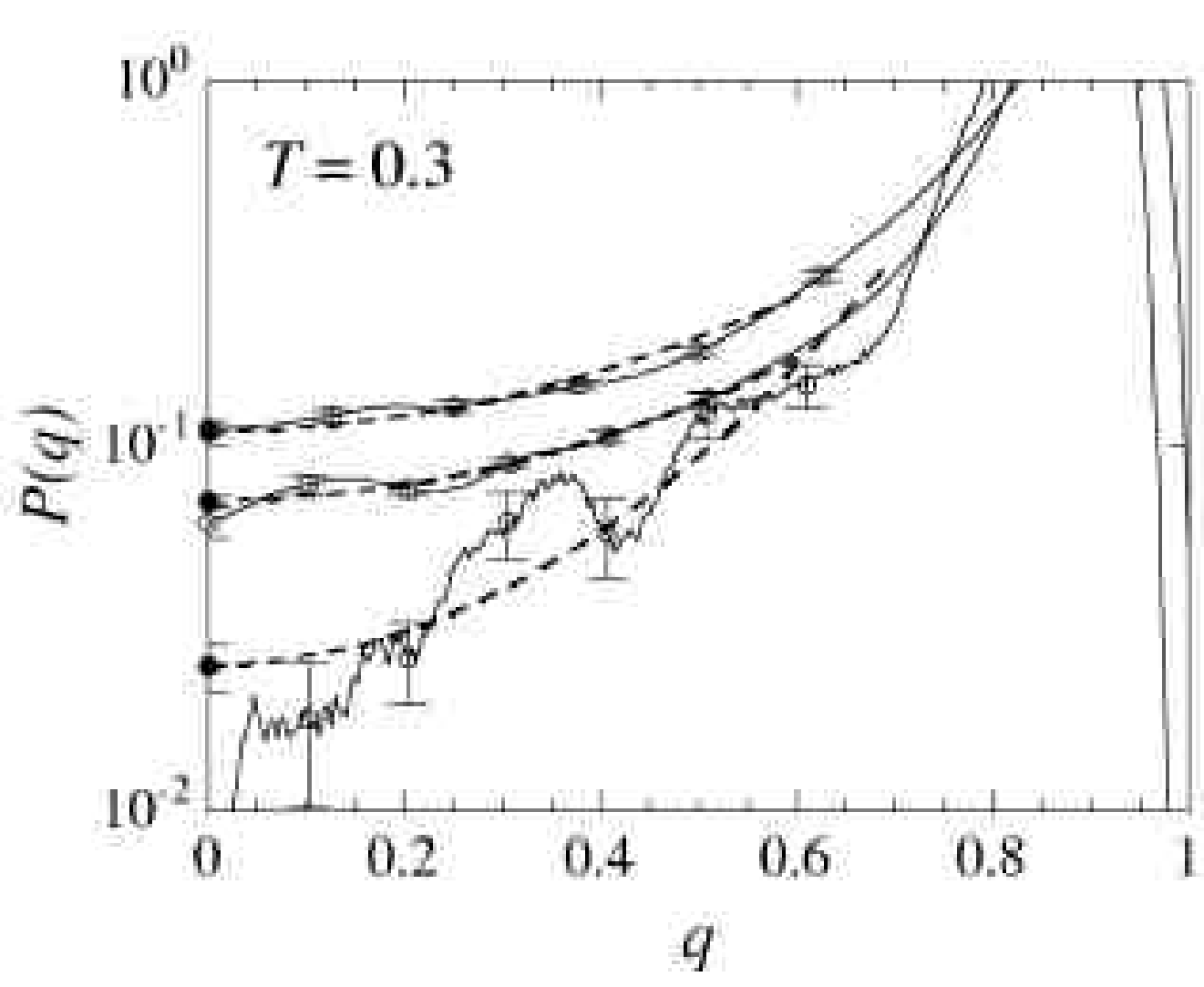}
\end{center}
\caption{The fitting curves for $P(q)$ and the estimates of
$P(0)$. Error bars are shown at only select points to illustrate the
typical error in different ranges of $q$.}
\label{fig-pofq-fit-0.3}
\end{figure}

To  obtain proper estimates
of $P(0)$, we have to smooth out the oscillations. We did this by
choosing seven data points at the intervals of $\Delta q\simeq0.1$ over
the range $0\leq q <0.7$ and then least-squares fitting them to the
function $\ln P(q)\simeq c_0+c_1q2$.
The fitting parameter $c_0$ yields the
estimate $P(0)=\exp(c_0)$.  The smoothed curves and the thus-estimated
values of $P(0)$ are shown in Fig.~\ref{fig-pofq-fit-0.3} for $T=0.3$.
We still see the decreasing tendency with increasing $L$.

\begin{figure}
\begin{minipage}[t]{0.35\textwidth}
\includegraphics[width=\textwidth]{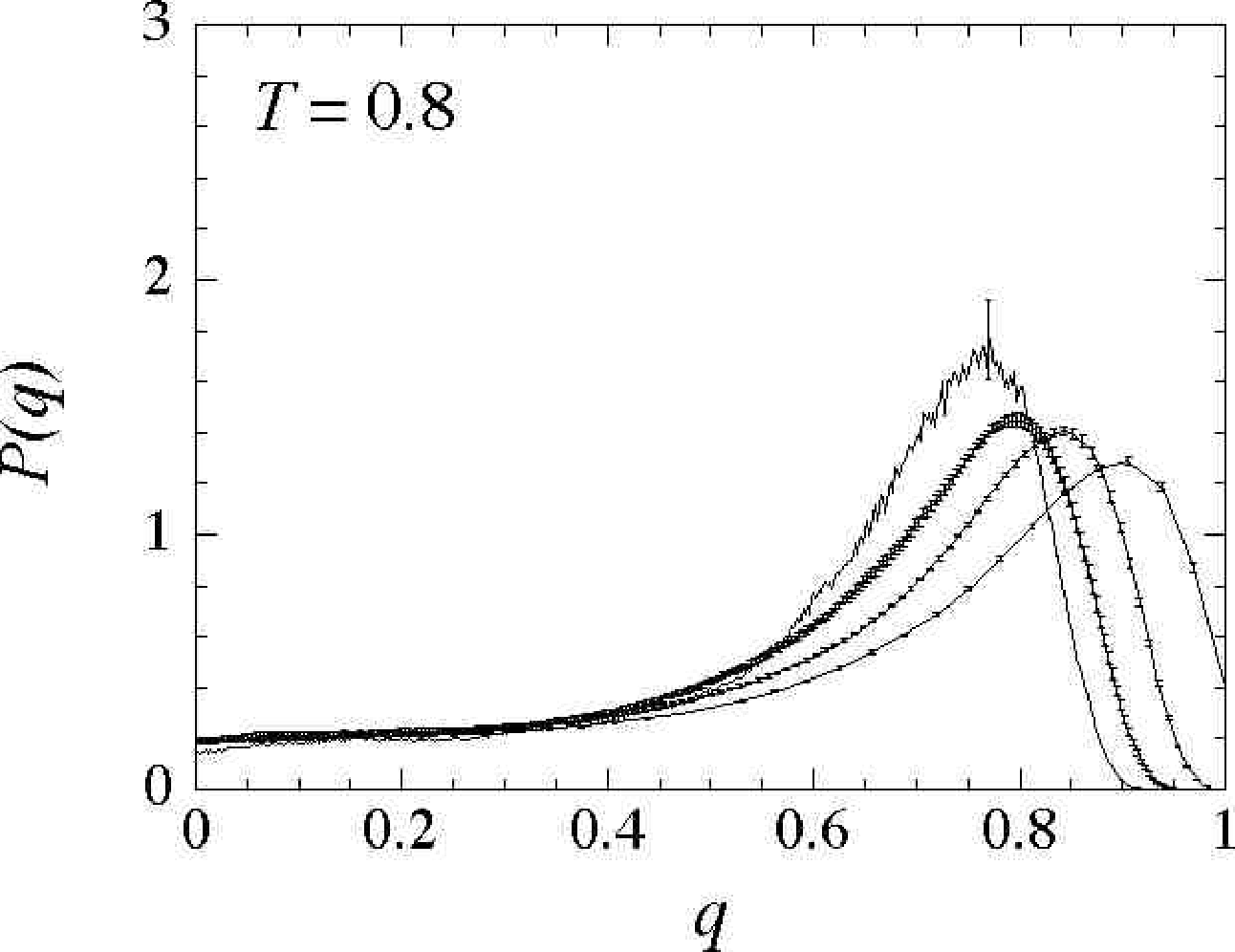}
(a)
\end{minipage}
\hfill
\begin{minipage}[t]{0.35\textwidth}
\includegraphics[width=\textwidth]{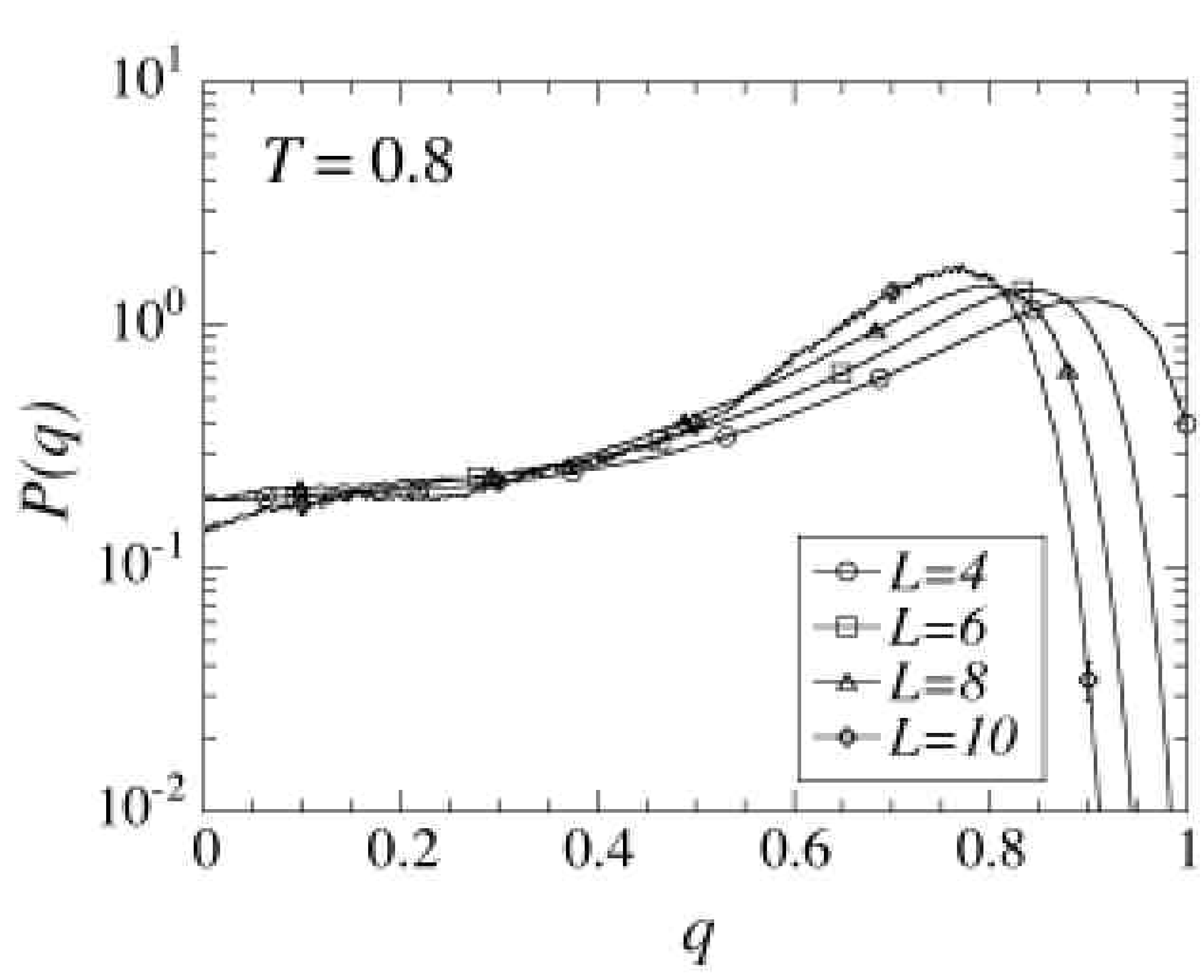}
(b)
\end{minipage}

\bigskip
\begin{minipage}[t]{0.35\textwidth}
\includegraphics[width=\textwidth]{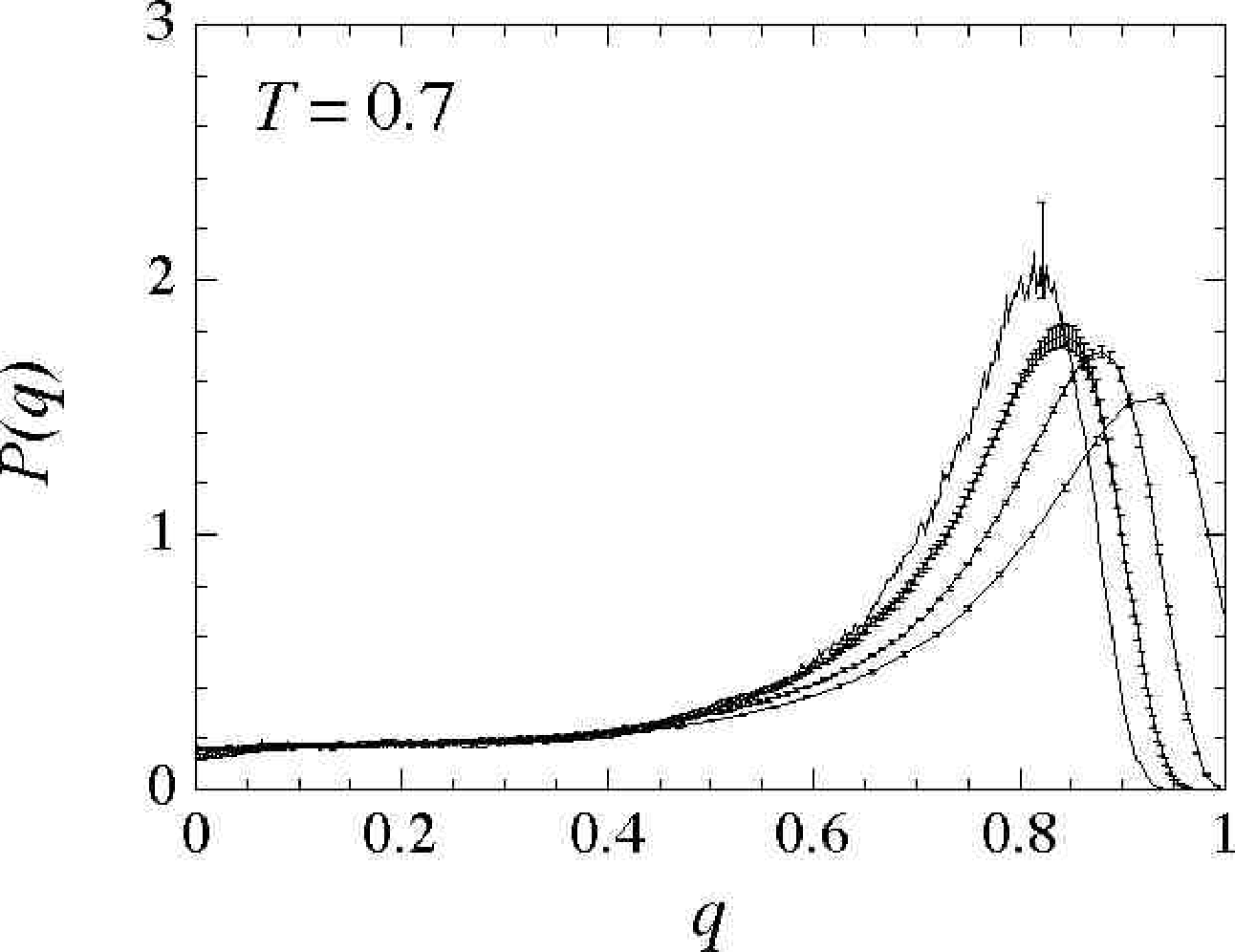}
(c)
\end{minipage}
\hfill
\begin{minipage}[t]{0.35\textwidth}
\includegraphics[width=\textwidth]{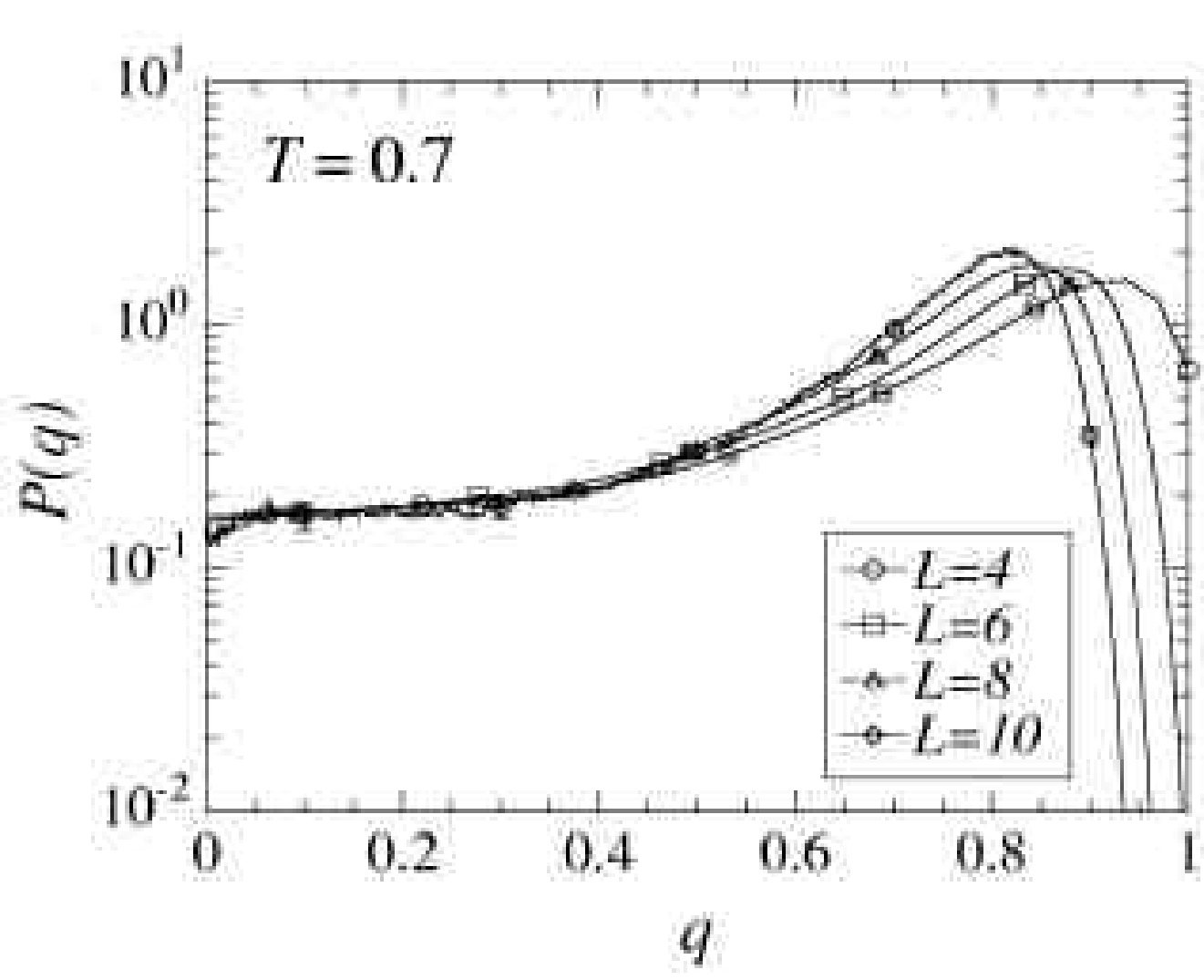}
(d)
\end{minipage}

\bigskip
\begin{minipage}[t]{0.35\textwidth}
\includegraphics[width=\textwidth]{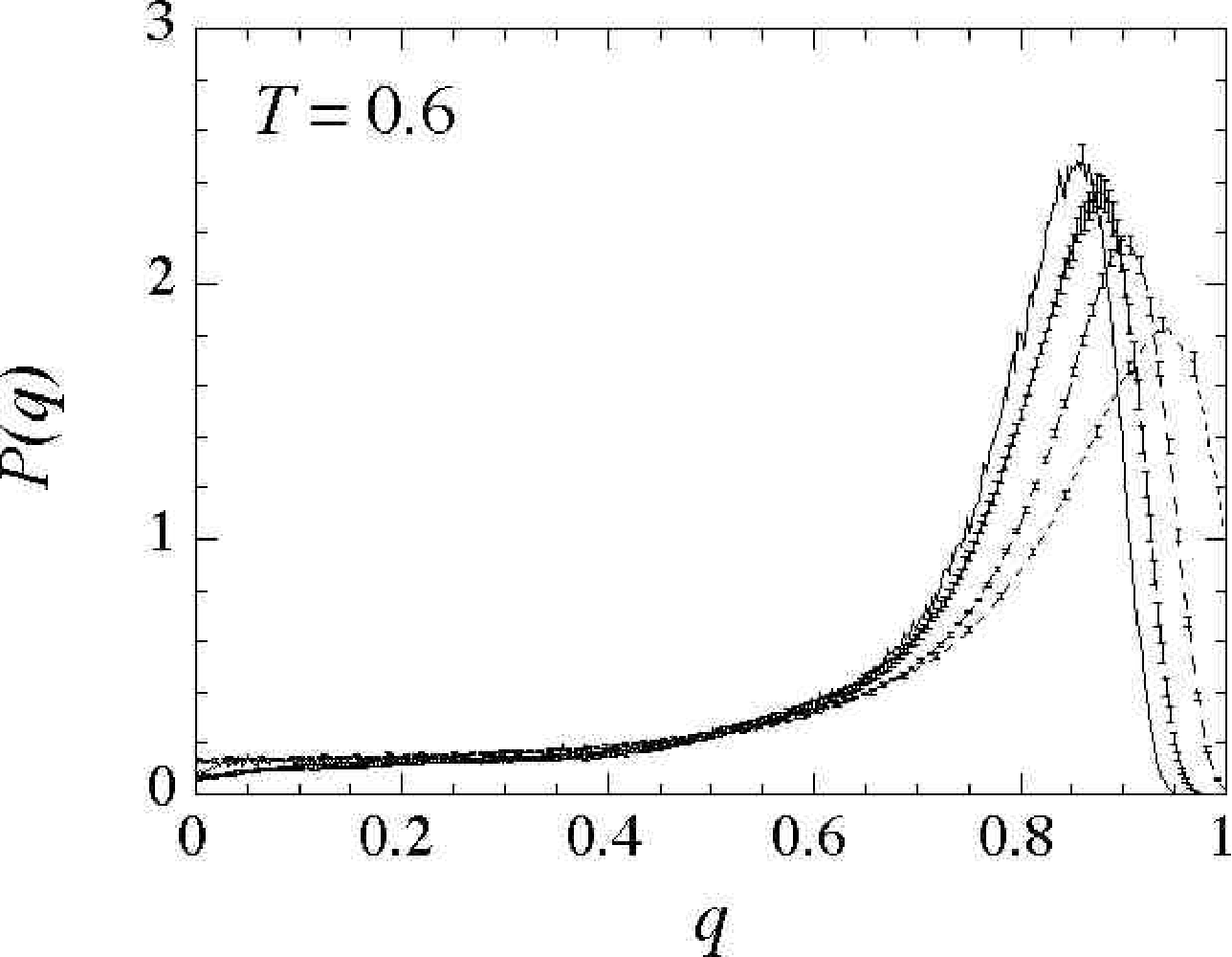}
(e)
\end{minipage}
\hfill
\begin{minipage}[t]{0.35\textwidth}
\includegraphics[width=\textwidth]{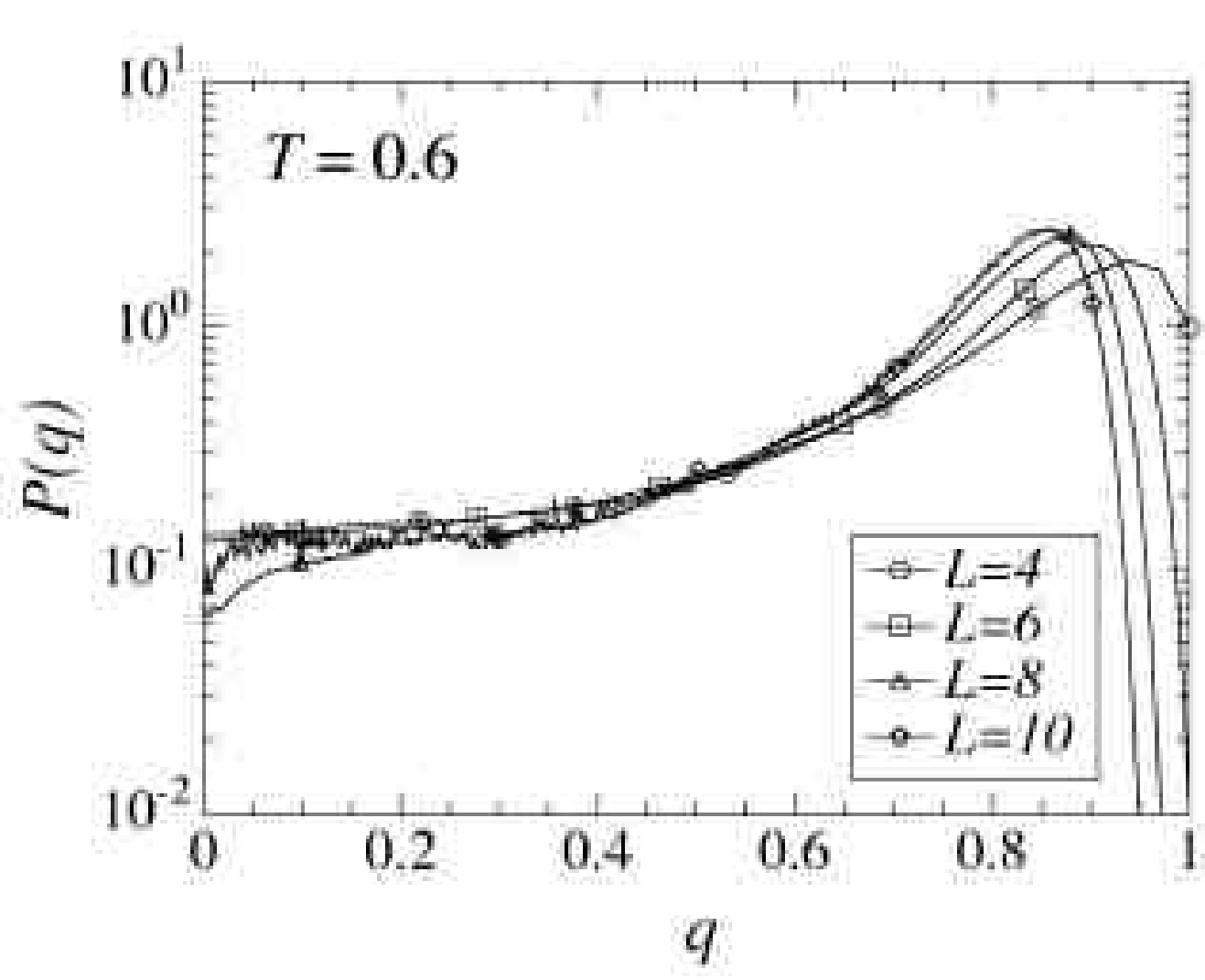}
(f)
\end{minipage}
\caption{
The size dependence of the order-parameter distribution $P(q)$.
(a) A linear plot for $T=0.8$ and $L=4,6,8$, and $10$.
The peak position moves left as the system size is increased.
Because the data points are very dense for $L=10$, the error bar is
shown only at the peak, where the statistical error is the largest.
(b) A semi-logarithmic plot for $T=0.8$.
The error bars are shown only for a part of the data points.
(c) A linear plot for $T=0.7$ and for $L=4,6,8$, and $10$.
The peak position moves left as the system size is increased.
For $L=10$, the error bar is shown only at the peak.
(d) A semi-logarithmic plot for $T=0.7$.
The error bars are shown only for a part of the data points.
(e) A linear plot for $T=0.6$ and for $L=4,6,8$, and $10$.
The peak position moves left as the system size is increased.
For $L=10$, the error bar is shown only at the peak.
(f) A semi-logarithmic plot for $T=0.6$.
The error bars are shown only for a part of the data points.
}
\label{fig-pofq-0.6-0.8}
\end{figure}

On the other hand, for $T\geq0.7$, $P(0)$ does not show this tendency,
even though its value is
calculated with the use of the same $D(E,q)$
(Fig.~\ref{fig-pofq-0.6-0.8}).
In fact, $P(0)$ appears to converge to a finite value, a behavior
which we now argue is spuriously consistent  with the
low-temperature behavior predicted by the mean-field picture.

\begin{figure}
\begin{center}
\includegraphics[width=8cm]{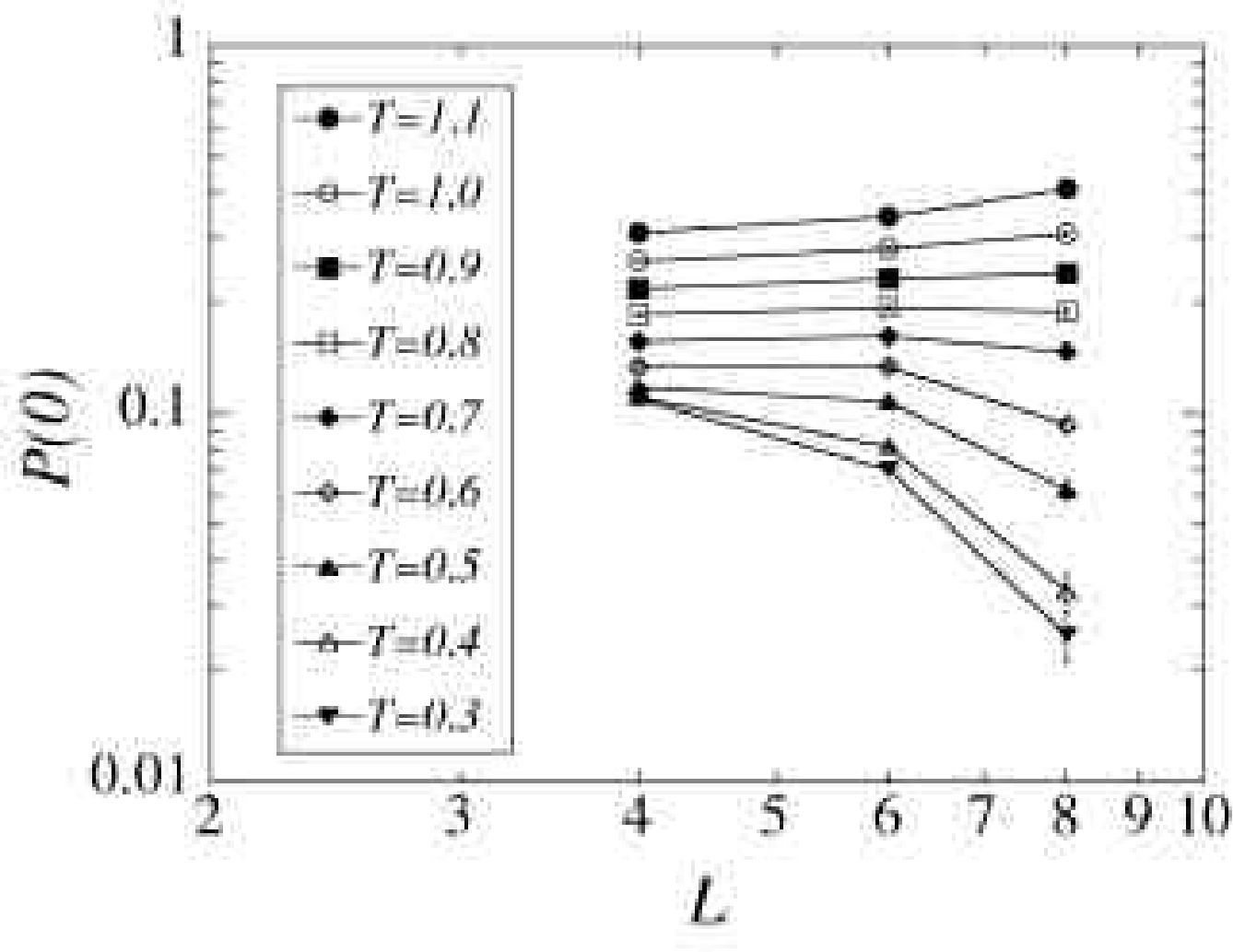}
\end{center}
\caption{The size dependence of $P(0)$ for
$T=1.1,1.0,0.9,\ldots,0.3$.  The data for $T=0.8$ {\em appear} to be
independent of $L$, but the data at lower temperatures reveal
features of the droplet picture.
The data in this figure were obtained after data processing
illustrated in Fig.~\protect\ref{fig-pofq-fit-0.3}.}
\label{fig-pof0}
\end{figure}

Independent of the degeneracy,
we know from the
scaling ansatz that at the critical point $T=T_{\mathrm{c}}\sim1$,
$P(0)$ should increase as $L\to\infty$.  On the other hand, at very
low temperatures, double degeneracy 
requires that $P(0)$ should decrease and infinite degeneracy
requires that it should tend to a constant.
What follows from our computed $D(E,q)$ for different system sizes is
shown in Fig.~\ref{fig-pof0}.  Near $T_c$, $P(0)$ does increase with
an increasing $L$. At lower temperatures, however, it tends to a
constant, a behavior supporting infinite degeneracy.
At still lower
temperatures, it tends to zero, a behavior consistent with double
degeneracy.
We note that a crossover scenario from critical, mean-field, to droplet
behavior was predicted from the Migdal-Kadanoff
approximation~\cite{Moore98}.  In particular we point out the
similarity of our Fig.~6 to their Fig.~5. We thus suggest that most
previous Monte Carlo studies, claiming to see behavior supporting the
mean-field picture (infinite degeneracy), based on results for $T$
only as low as $0.7$, missed behavior consistent with the droplet
picture (double degeneracy) which only appears at much lower
temperatures.

Palassini and Young \cite{Palassini01} have studied the scaling of the
smoothed quantity $x(1/2)=\int_{-1/2}^{1/2}P(q)dq$, and found a
crossover scaling between $T=0$ behavior, where $x(1/2)$ becomes
trivial for $L\rightarrow \infty$ and finite-temperature behavior,
where the non-trivial part of $P(q)$ has a much weaker dependence on
$L$ and is possibly size independent. The crossover is consistent with
the qualitative features of our results. In fact, Palassini and
Young's Fig.~4, showing this crossover, is strikingly similar to our
Fig.~\ref{fig-pof0}.  The remark by Palassini and Young that our
$P(0)$ drops dramatically at low $T$ as $L$ increase was based on the
preliminary analysis of our data where $P(0)$ was determined in the
absence of the smoothing. Palassini and Young also point to possible
different behavior between models with Gaussian distributed exchange
interactions, where $P(0)$ might be non-trivial as
$L\rightarrow\infty$ and bimodal ones, where $P(0)$ becomes trivial.

We also note the recent zero temperature work by Hed, Domany, and
Hartmann~\cite{Hed01}. These investigators also found the need to
smooth the values of $P(q)$, and for better statistics they chose to
study $x^*=\int_{0.4}^{0.7}P(q) dq$. In the ground state they claim
that $x^*$ scales to a small non-trivial value. To be more precise,
they first separated $P(q)$ into a part that comes from the big peaks
close to $q=\pm 1$ that have an $L$-dependent tail at $q=0$ which
scales to zero and into a part more proper to ground-state excitations
whose scaling with $L$ is the central issue.  For this latter part
they claim $x^*$ is non-trivial in the thermodynamic limit.

\subsection{Binder parameter}

The Binder parameter for spin glasses $g_{\mathrm{sg}}$, defined by
Eq.~(\ref{eq-binder}), is essentially the kurtosis of
the order-parameter distribution $P(q)$.  Because of the dimensionless
combination of the second moment and the fourth moment, this
parameter (except for effects due to correction to scaling) is
expected to be independent of the system size at fixed points, {\it
i.e.\/}, $T=0$, $T=T_{\mathrm{c}}$, and $T\rightarrow \infty$ .  For
conventional phase transitions such as ferromagnetic transitions, its
temperature dependence for various system
sizes has a crossing point at the critical temperature.  At the
high-temperature $(T\rightarrow\infty)$ fixed point, the order-parameter
distribution $P(q)$ should be Gaussian.  Hence we should have
\begin{equation}
\int q4 P(q)dq=3\left(\int q2 P(q)dq\right)2,
\end{equation}
or
\begin{equation}
\left[\left\langle q4\right\rangle\right]_{\mathrm{av.}}=
3\left[\left\langle q2\right\rangle\right]_{\mathrm{av.}}^2.
\end{equation}
Thus the high temperature fixed-point value of the Binder parameter is
zero, {\it i.e.\/}, $g_{\mathrm{sg}}(\infty,L)=0$.  In the
high-temperature phase, the Binder parameter $g_{\mathrm{sg}}(T,L)$ is
renormalized to be zero from above as $L\to\infty$.  At the
low-temperature ($T=0$) fixed point, on the other hand, if the order
parameter takes only two values $\pm q_{0}$, as in the thermodynamic
limit of usual ferromagnets, we have
\begin{equation}
\left[\left\langle q^{4} \right\rangle\right]_{\mathrm{av.}}
=\left[\left\langle q^{2} \right\rangle\right]_{\mathrm{av.}}^2=q_{0}^{4}
\end{equation}
and hence the fixed-point value of the Binder parameter is
unity: $g_{\mathrm{sg}}(0,L)=1$.  In the low-temperature phase, the
Binder parameter $g_{\mathrm{sg}}(T,L)$ is renormalized to
be unity from below as $L\to\infty$.  At $T=T_{\mathrm{c}}$, the Binder
parameter $g_{\mathrm{sg}}(T_{\mathrm{c}},L)$ is expected to have a
nontrivial universal value between zero and unity.  Thus the crossing
point of $g_{\mathrm{sg}}(T,L)$ should give the critical temperature
$T_{\mathrm{c}}$.

\begin{figure}
\begin{center}
\includegraphics[width=8cm]{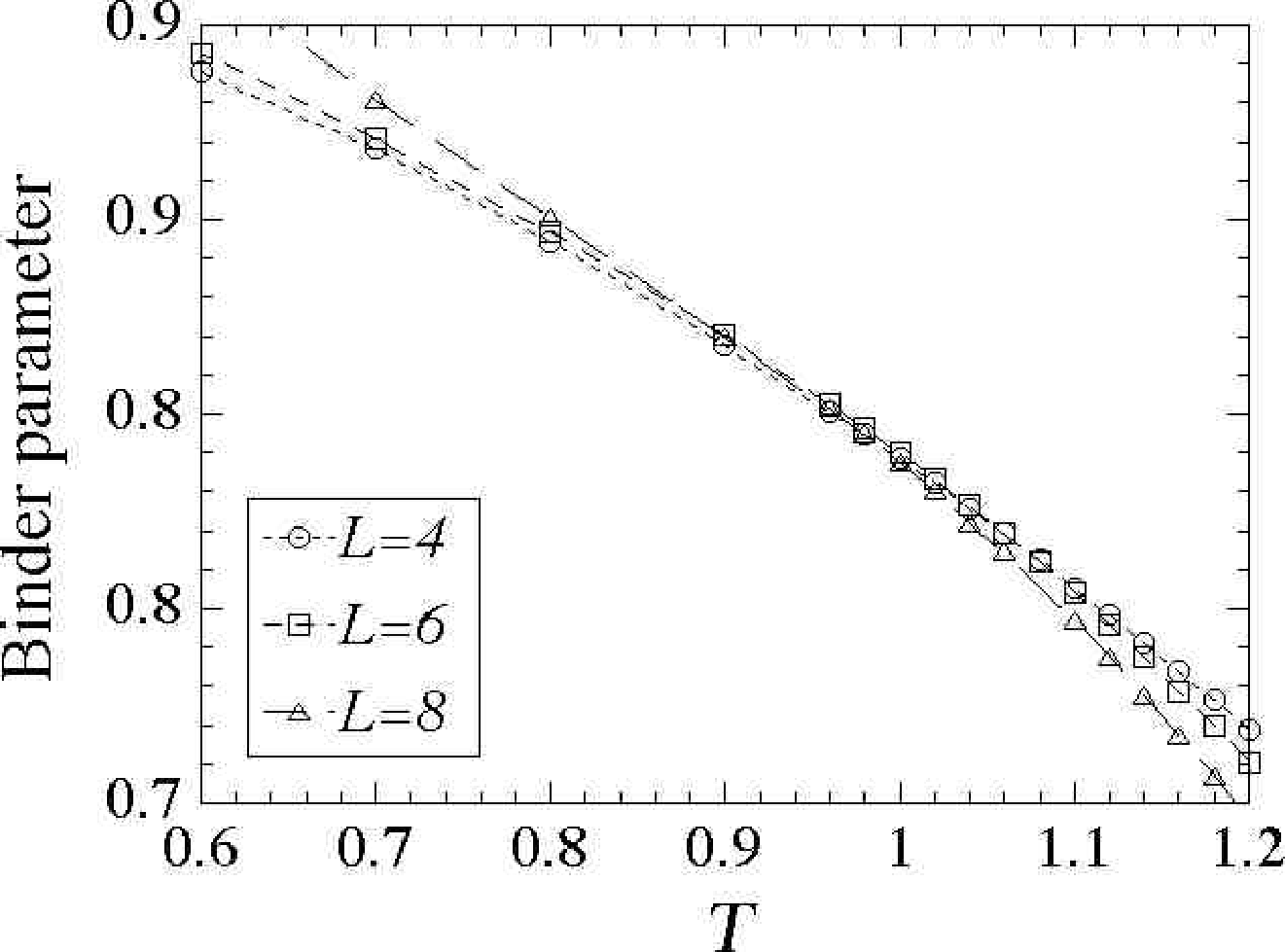}
\end{center}
\caption{The temperature dependence of the Binder parameter $g(T,L)$
for $L=4,6$, and $8$.  The statistical errors are comparable to the symbol
size.}
\label{fig-binder}
\end{figure}

Our Monte Carlo simulation found the crossing point of the Binder
parameter as shown in Fig.~\ref{fig-binder} \cite{note2}.
The critical point $T_c$ should be in the region $0.8\geq
T_{\mathrm{c}}\geq1.1$.  For the moment, because of strong corrections
to scaling, it is difficult for us to carry out
sophisticated scaling analysis and obtain a more accurate estimate of
$T_{\mathrm{c}}$.  Previous
studies~\cite{Kawashima,Hukushima,pal_and_car,Ballesteros} on much 
larger systems claim
$T_{\mathrm{c}}\simeq1.1$.

\begin{figure}
\begin{center}
\includegraphics[width=8cm]{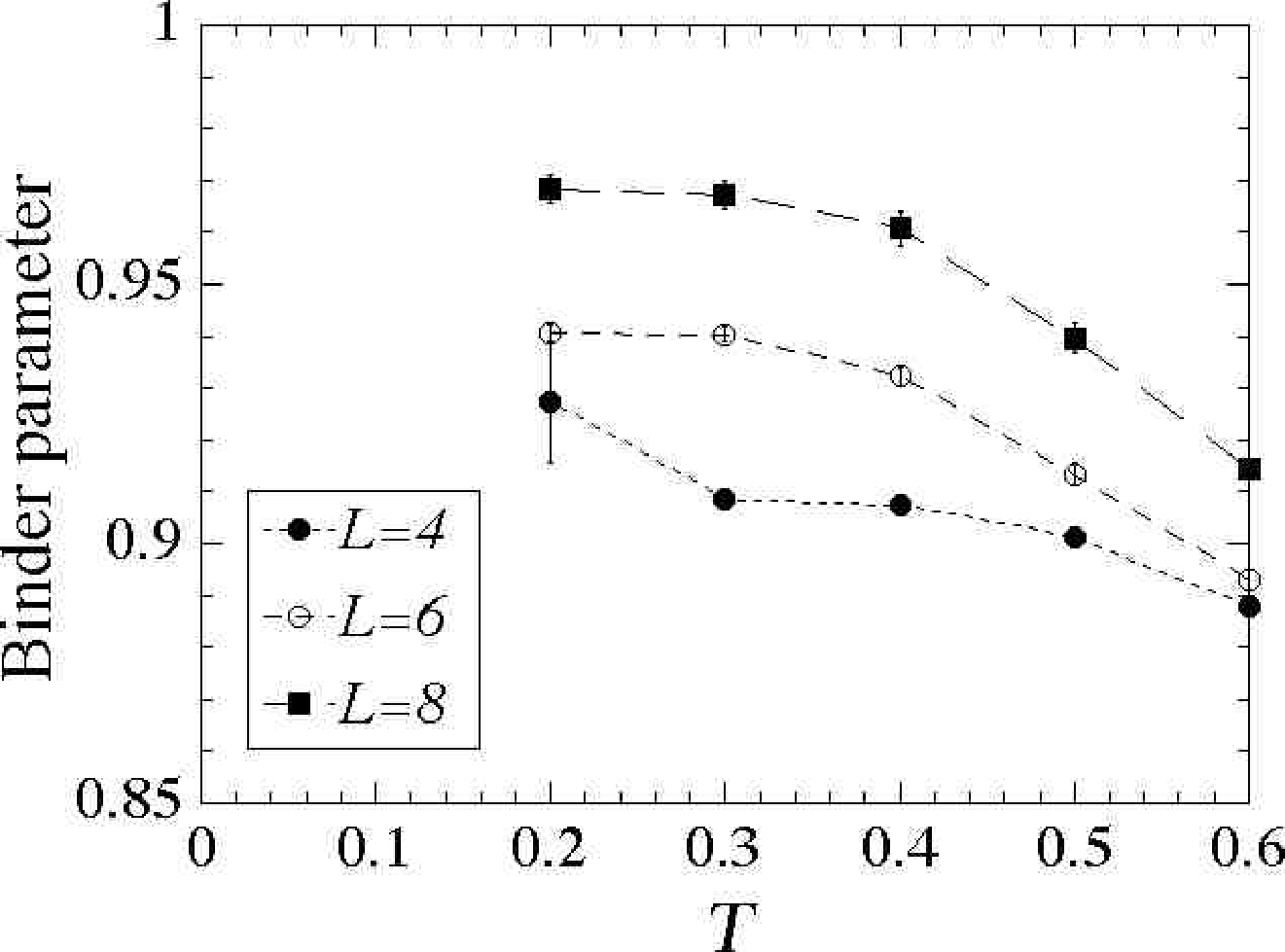}
\end{center}
\caption{The temperature dependence of the Binder parameter $g(T,L)$
for $L=4,6$, and $8$.}
\label{fig-binder0}
\end{figure}

\begin{figure}
\begin{center}
\includegraphics[width=8cm]{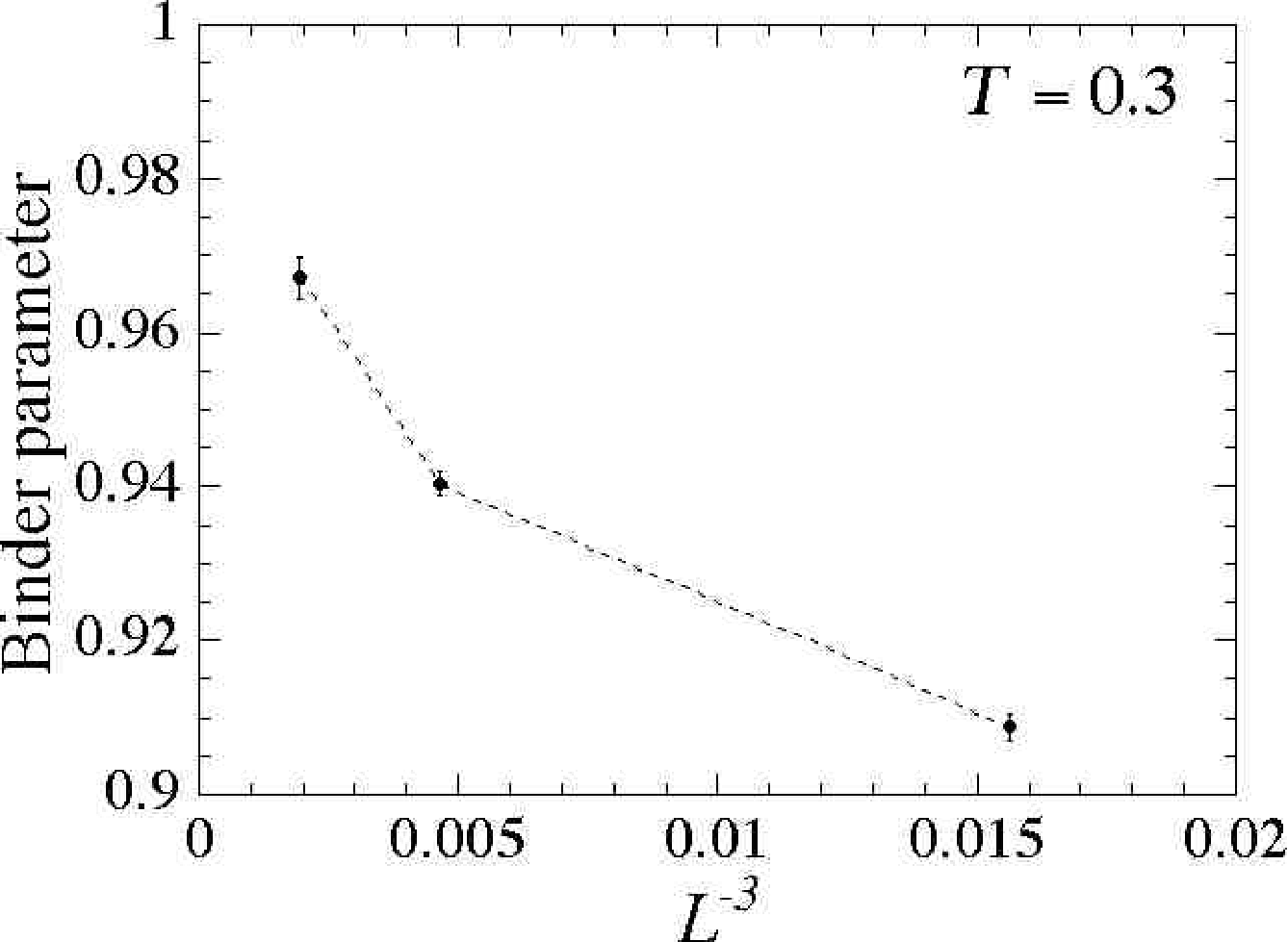}
\end{center}
\caption{The size dependence of the Binder parameter at $T=0.3$.
The Binder parameter is approaching unity rapidly.}
\label{fig-binderT0.3}
\end{figure}

We now offer further evidence for a doubly degenerate ground state, as
opposed to an infinitely degenerate one, by reporting our results for
the low-temperature behavior of the Binder parameter in
Fig.~\ref{fig-binder0}.
These results strongly suggest that the Binder parameter tends to unity
as $T\to0$ and $L\to\infty$.
This behavior is consistent with 
Fig.~\ref{fig-pofq-schem}a and not with 
Fig.~\ref{fig-pofq-schem}b, as the Binder parameter is less than unity
even at $T=0$ and $L\to\infty$.
Our Monte Carlo results thus clearly
support double degeneracy (see Fig.~\ref{fig-binderT0.3}).

\subsection{Residual entropy}
\begin{figure}
\begin{minipage}{0.45\textwidth}
\includegraphics[width=\textwidth]{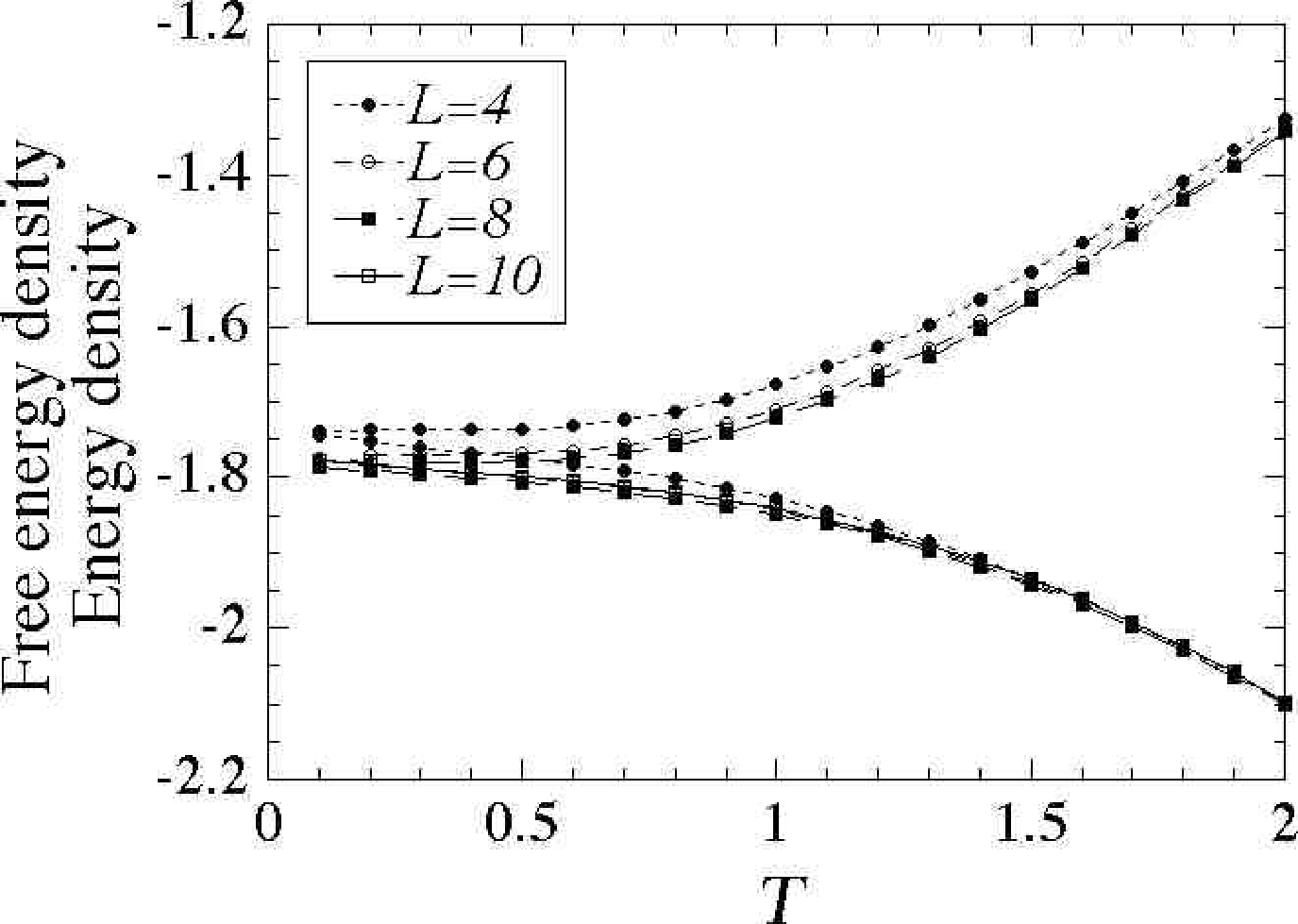}
(a)
\end{minipage}
\hfill
\begin{minipage}{0.45\textwidth}
\includegraphics[width=\textwidth]{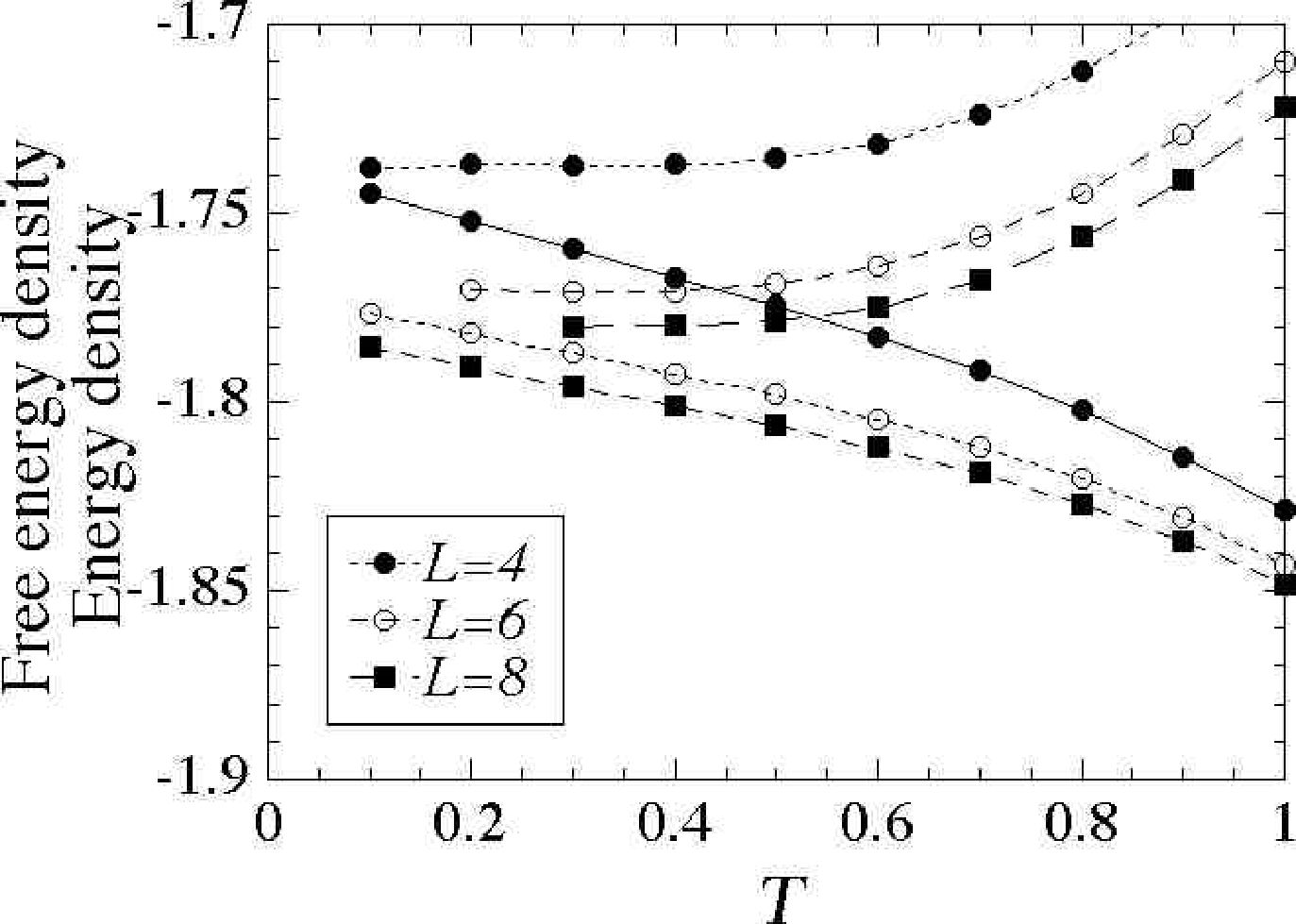}
(b)
\end{minipage}
\caption{The temperature dependence of the energy density and the
free-energy density. (b) is an enhanced view of (a). The statistical
errors are smaller or comparable to the symbol size.}
\label{fig-eandf}
\end{figure}

\begin{figure}
\begin{minipage}{0.45\textwidth}
\includegraphics[width=\textwidth]{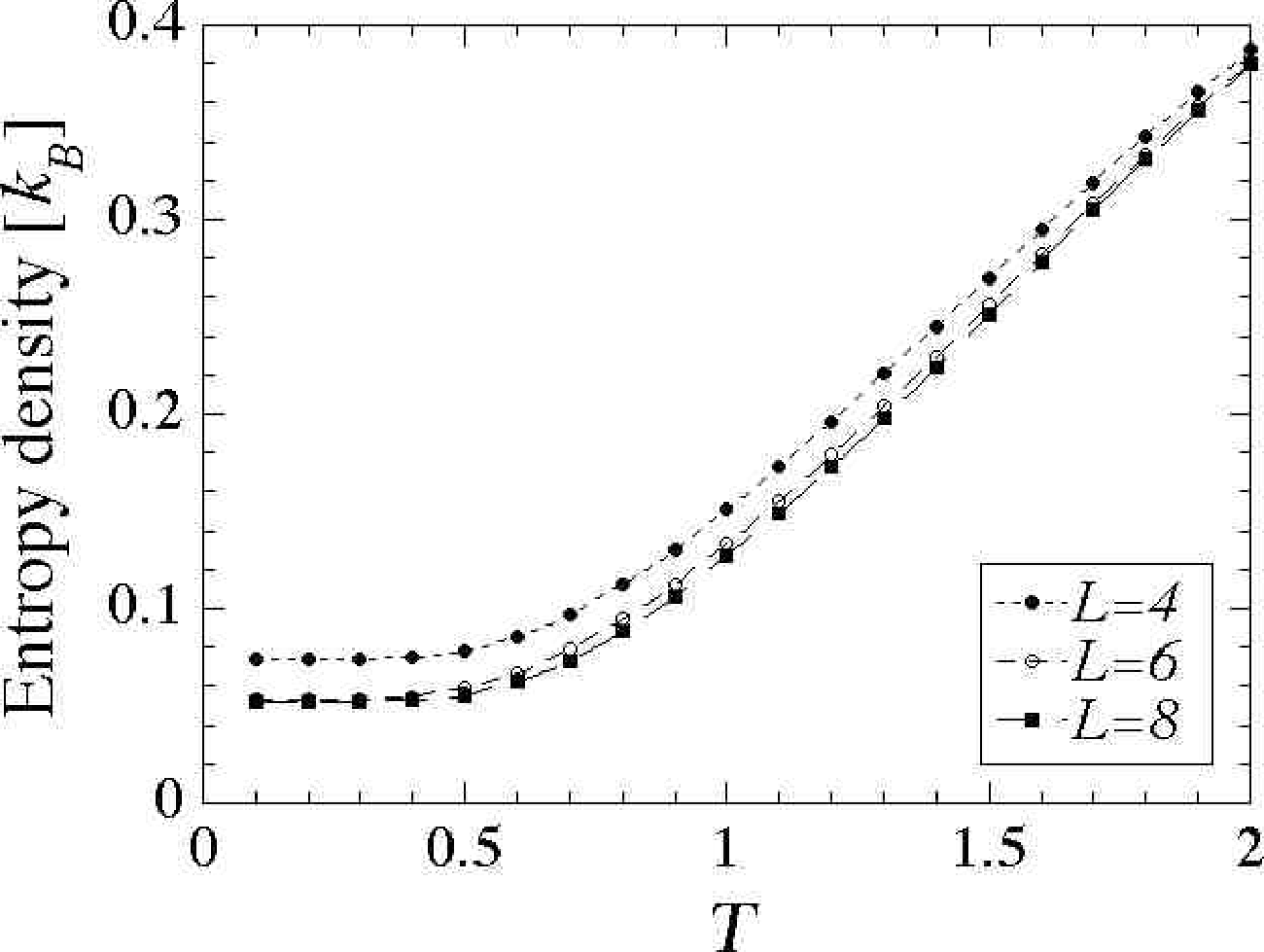}
(a)
\end{minipage}
\hfill
\begin{minipage}{0.45\textwidth}
\includegraphics[width=\textwidth]{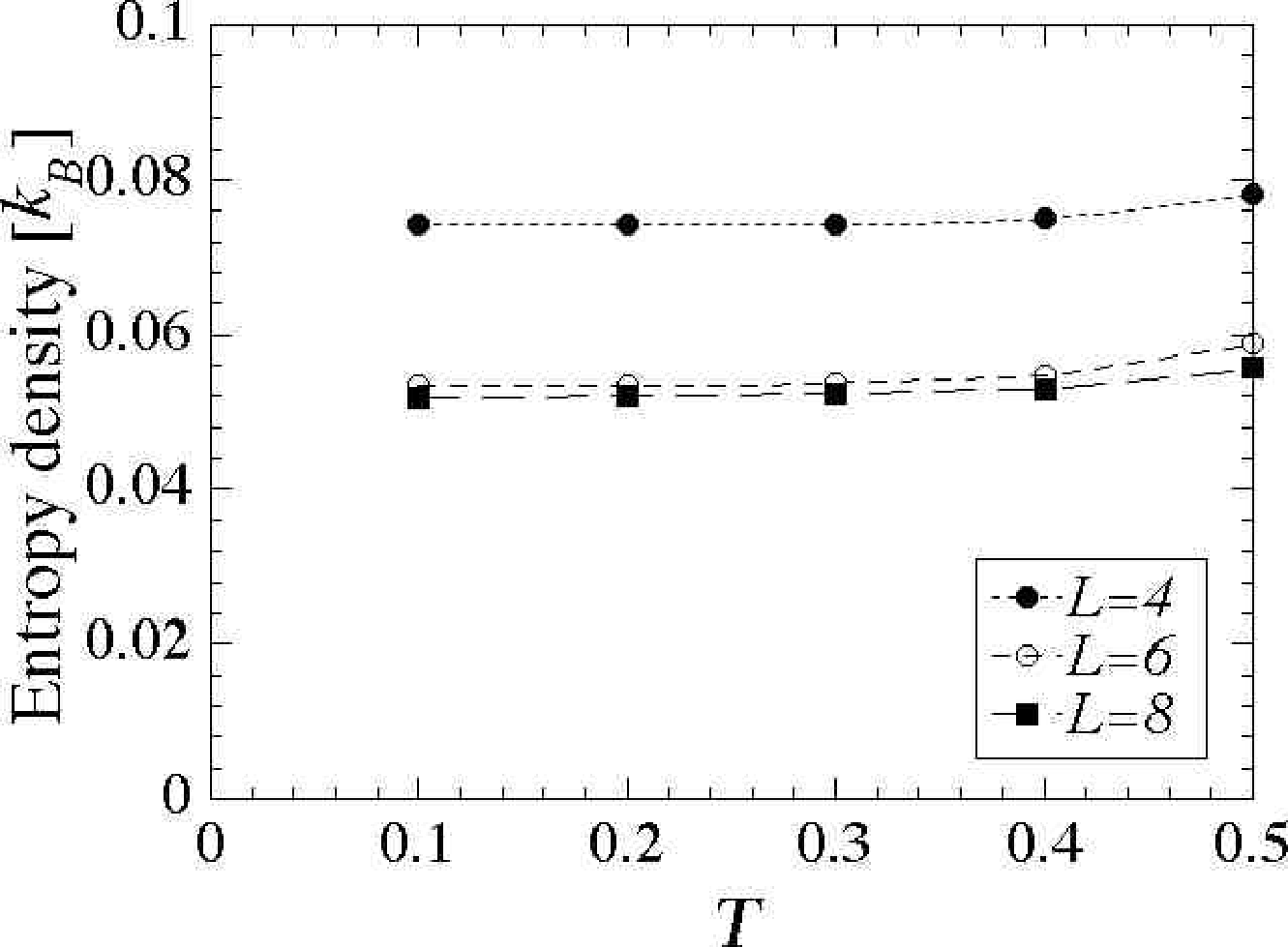}
(b)
\end{minipage}
\caption{The temperature dependence of the entropy density. (b) is an
enhanced view of (a). The statistical errors are smaller or comparable
to the symbol size.}
\label{fig-entropy}
\end{figure}

\begin{figure}
\begin{center}
\includegraphics[width=8cm]{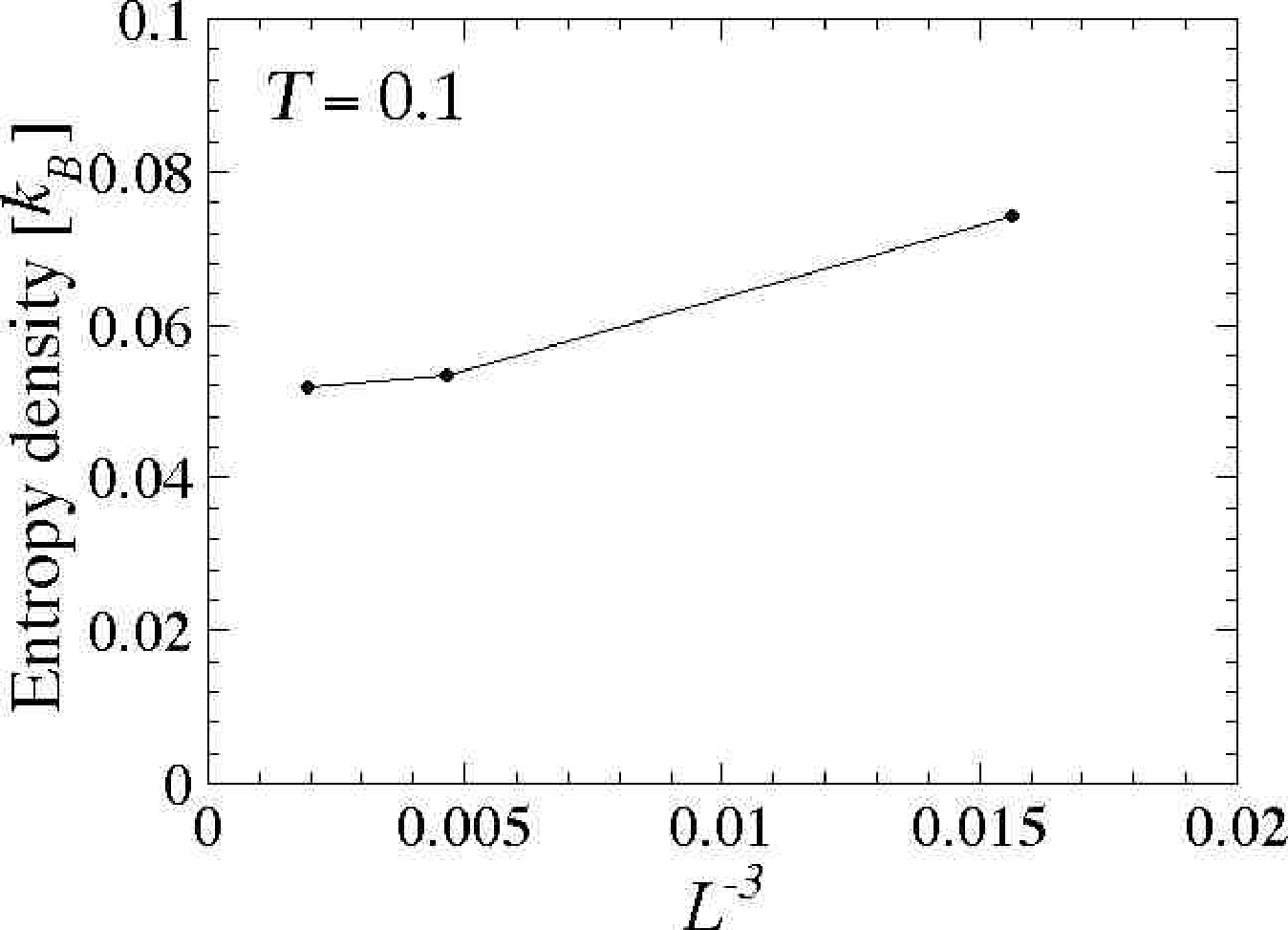}
\end{center}
\caption{The size dependence of the residual entropy. The statistical
errors are smaller or comparable to the symbol size.}
\label{fig-sT0.1}
\end{figure}

We calculated the entropy density from the difference between the
energy and the free energy, {\it i.e.}
\begin{equation}
s=\frac{\left[\left\langle E \right\rangle\right]_{\mathrm{av.}}-F}{NT},
\end{equation}
where $F$ is the free energy calculated from the Monte Carlo output
$D(E,q)$ via
\begin{equation}
F=-\frac{1}{\beta}\log Z
=-\frac{1}{\beta}\log \left(\sum_{E,q}D(E,q)e^{-\beta E}\right).
\end{equation}
Figure~\ref{fig-eandf} shows the energy and free-energy
densities.
The difference between them is shown in Fig.~\ref{fig-entropy}.
It is clear that at $T=0.1$ the estimates of the entropy are virtually
the residual entropy at $T=0$. The residual entropy is
plotted in Fig.~\ref{fig-sT0.1}.
The convergence to the thermodynamic limit is quite rapid and
the entropy seems to remain finite in the thermodynamic limit.
Hartmann~\cite{Hartmann99} also observed a rapid convergence to a
finite entropy from a ground-state search on systems up
to $L=8$.
He obtained a similar estimate of the residual entropy
$s(T=0)=0.051(3)k_{\mathrm{B}}$.

Although Hartmann used the existence of the
residual entropy as evidence for the mean-field picture, its existence
is on the contrary entirely consistent with the droplet picture.  The
degeneracy of the ground states predicted by these
pictures
is the degeneracy of thermodynamic (macroscopic)
states, while the residual entropy comes from the degeneracy of
microscopic states. The distinction is important to note.

\begin{figure}
\begin{center}
\includegraphics[width=8cm]{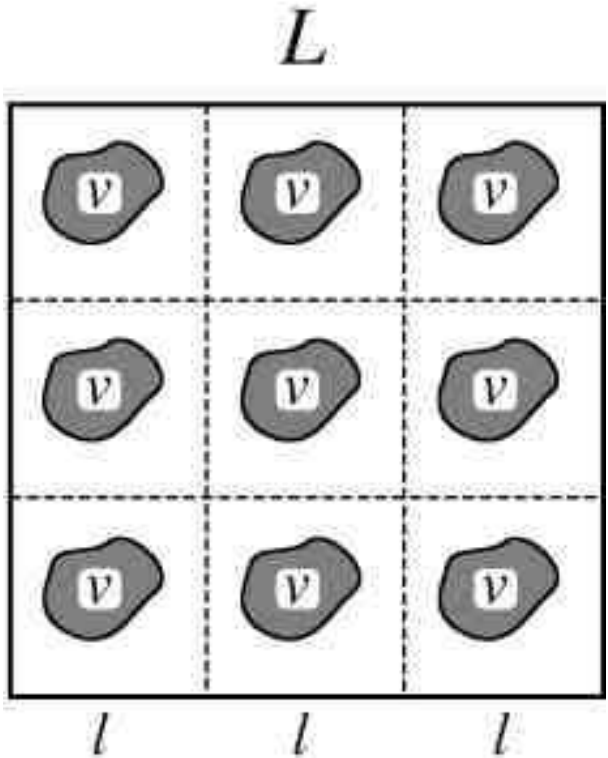}
\end{center}
\caption{Toy model with a finite residual entropy density and
delta-function peaks in $P(q)$. In the figure, $L$ is the linear size
of the whole system, $l$ is the size of the unit cell, and $v$ is the
volume of a zero-energy droplet.}
\label{fig-toymodel}
\end{figure}

Because the energy of a finite-sized $\pm J$ model is discrete, there is an
inevitable degeneracy of the ground states.  The issue is whether the
degeneracy arises from these microscopically degenerate states
or from many macroscopically
different states.  To make the distinction clearer, we will now
consider a toy model in
which we quench the exchange interactions into a periodic
configuration with a unit cell of linear size $l$.  This model has a
ground state with a periodic spin configuration.  We will assume,
however, that every unit cell has one connected cluster of spins such
that the spin inversion of the cluster does not change the
ground-state energy (See Fig.~\ref{fig-toymodel}).
We refer to such spin clusters as ``zero-energy droplets"
\cite{bray,barahona}. The number
of such droplets in a system of linear size $L$ is
$N_{\mathrm{zed}}=(L/l)^d$, where $d$ is the dimensionality.  The
degeneracy of the ground-state energy is $2^{N_{\mathrm{zed}}}$.
Therefore, the residual entropy density of the toy model takes a
finite value
\begin{equation}
s_{\mathrm{toy}}(T=0)=
L^{-d}\times
k_{\mathrm{B}}
\ln \left(
2^{N_{\mathrm{zed}}}
\right)
=k_{\mathrm{B}}
l^{-d}\ln2.
\end{equation}

This model, on the other hand, produces a $P(q)$ consistent with
the droplet picture: Consider two replicas of the toy model.  Without
loss of generality, we can fix the spin configuration of one replica
and calculate contributions from different spin configurations of the
other replica.  The distribution of the order parameter is given by
\begin{equation}\label{eq-toy-dist}
\begin{array}{lll}
|q|=1 &
\mbox{with the probability} &
2^{-N_{\mathrm{zed}}},
\\
|q|=1-2v/L^d &
\mbox{with the probability} &
2^{-N_{\mathrm{zed}}}\times N_{\mathrm{zed}},
\\
|q|=1-4v/L^d &
\mbox{with the probability} &
2^{-N_{\mathrm{zed}}}\times \frac{1}{2}N_{\mathrm{zed}}(N_{\mathrm{zed}}-1),
\\
\cdots & & \\
|q|=1-2nv/L^d &
\mbox{with the probability} &
2^{-N_{\mathrm{zed}}}\times
\left(\begin{array}{c}
N_{\mathrm{zed}} \\ n
\end{array}\right)
\\
\cdots, & &
\end{array}
\end{equation}
where $v$ is the volume of each zero-energy droplet.  In the
thermodynamic limit $L\to\infty$ or equivalently
$N_{\mathrm{zed}}\to\infty$, the
probability for $n=N_{\mathrm{zed}}/2$ overwhelms and hence
$P(q)$ converges to a delta-function peak at
\begin{equation}
q=1-\frac{N_{\mathrm{zed}}}{2}{2v}{L^d}=1-\frac{v}{l^d}.
\end{equation}
and to a second delta-function peak at $-q$.

\section{Ground-state degeneracy of the $\pm J$ model}\label{sec3}

\begin{figure}
\begin{center}
\includegraphics[width=8cm]{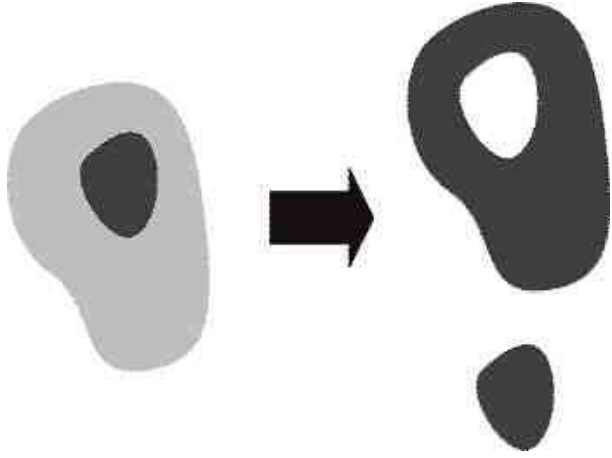}
\end{center}
\caption{If a zero-energy droplet contains another zero-energy droplet
as shown on the left, we define the volumes of the two droplets as
shown on the right.}
\label{fig-zed}
\end{figure}

The toy model just introduced
is easily converted to a more realistic one for the ground-state
degeneracy of the $\pm J$ model
by allowing
unequally-sized droplet volumes $v_i$
($i=1,2,3,\ldots,N_{\mathrm{zed}}(L)$).
In defining the volume of these droplets, we always take the
minimal volume; for example, if a zero-energy droplet contains another
zero-energy droplet, we decompose them into two droplets as shown in
Fig.~\ref{fig-zed}.
In addition, the possible maximum volume of the zero-energy droplet is
$L^d/2$: if there is a zero-energy droplet larger than the half of the
system, we re-define the original ground-state spin configuration by
flipping the largest zero-energy droplet.

The ground-state degeneracy is $2^{N_{\mathrm{zed}}(L)}$, and hence
the residual entropy density is given by
\begin{equation}\label{eq-entropy1}
s(T=0;L)=
L^{-d}
k_{\mathrm{B}}
\ln \left(
2^{N_{\mathrm{zed}}(L)}
\right)
=L^{-d}
N_{\mathrm{zed}}(L)
k_{\mathrm{B}}
\ln 2.
\end{equation}
We now define the distribution of the zero-energy droplets as
$n_{\mathrm{zed}}(v)$ so that we have
\begin{equation}
N_{\mathrm{zed}}(L)=L^d\int_1^{L^d/2}n_{\mathrm{zed}}(v)dv.
\end{equation}
and rewrite Eq.~(\ref{eq-entropy1}) as
\begin{equation}\label{eq-entropy10}
s(T=0;L)=k_{\mathrm{B}}\ln 2\times
\int_1^{L^d/2}n_{\mathrm{zed}}(v)dv.
\end{equation}

Next, we consider the ground-state value of $\langle |q| \rangle$, and
again we fix the spin configuration of one replica to the original
ground-state spin configuration.  Various contributions come from the
spin configurations of the other replicas with some of the zero-energy
droplets flipped.  We extend Eq.~(\ref{eq-toy-dist}) and obtain
\begin{equation}\label{eq-q-dist}
\begin{array}{ll}
|q|=1 &
\mbox{with the probability $2^{-N_{\mathrm{zed}}}$,}
\\
|q|=1-2v_i/L^d &
\mbox{with the probability $2^{-N_{\mathrm{zed}}}$ for
$i=1,2,\ldots,N_{\mathrm{zed}}$,}
\\
|q|=1-2(v_i+v_j)/L^d &
\mbox{with the probability $2^{-N_{\mathrm{zed}}}$ for
$i,j=1,2,\ldots,N_{\mathrm{zed}}$
with $i<j$,}
\\
\cdots & \\
|q|=1-(2/L^d) \sum_{m=1}^n v_{i_m} &
\mbox{with the probability $2^{-N_{\mathrm{zed}}}$} \\
&\mbox{for $i_1,i_2,\ldots,i_n=1,2,\ldots,N_{\mathrm{zed}}$ with
$i_1<i_2<\cdots<i_n$,}
\\
\cdots &
\end{array}
\end{equation}
Summing all contributions, we find
\begin{equation}
\langle |q| \rangle
=1
-\frac{2}{2^{N_{\mathrm{zed}}}L^d}
\left[
\sum_{i}v_i
+\sum_{i < j}(v_i+v_j)
+\cdots\right]
=1-\alpha_{\mathrm{zed}},
\end{equation}
where
\begin{equation}
\alpha_{\mathrm{zed}}(L)\equiv
\frac{1}{L^d}\sum_{i=1}^{N_{\mathrm{zed}}}v_i
=\int_1^{L^d/2} v n_{\mathrm{zed}}(v)dv.
\end{equation}

We can now compute $\langle q2 \rangle$ in the same way and obtain
\begin{equation}
\langle q2 \rangle
=
\frac{1}{2^{N_{\mathrm{zed}}}}
\left[
\sum_{i}\left\{1-\frac{2}{L^d}v_i\right\}^2
+\sum_{i < j}\left\{1-\frac{2}{L^d}(v_i+v_j)\right\}^2
+\cdots\right].
\end{equation}
which reduces to
\begin{equation}
\langle q2 \rangle=(1-\alpha_{\mathrm{zed}})2
+\beta_{\mathrm{zed}},
\end{equation}
or
\begin{equation}\label{eq-beta}
\langle q2 \rangle-\langle |q| \rangle2=\beta_{\mathrm{zed}},
\end{equation}
where
\begin{equation}\label{eq-betadef}
\beta_{\mathrm{zed}}(L)\equiv
\frac{1}{L^{2d}}\sum_{i=1}^{N_{\mathrm{zed}}}v_i2
=\frac{1}{L^{d}}\int_1^{L^d/2} v2 n_{\mathrm{zed}}(v)dv.
\end{equation}
Similarly we have
\begin{equation}\label{eq-gamma}
\langle q4 \rangle -3 \langle q2 \rangle2 +2 \langle |q| \rangle4
=-2\gamma_{\mathrm{zed}},
\end{equation}
where
\begin{equation}\label{eq-gammadef}
\gamma_{\mathrm{zed}}(L)\equiv
\frac{1}{L^{4d}}\sum_{i=1}^{N_{\mathrm{zed}}}v_i4
=\frac{1}{L^{3d}}\int_1^{L^d/2} v4 n_{\mathrm{zed}}(v)dv.
\end{equation}

If
the order-parameter distribution
$P(q)$ has only two delta-function peaks at the $L\to\infty$ limit,
then the left-hand sides of Eq.~(\ref{eq-beta}) and
Eq.~(\ref{eq-gamma}) vanish.  This condition is satisfied
if the density distribution of the zero-energy droplets
$n_{\mathrm{zed}}(v)$ decays fast enough as $v\to\infty$.  If
$n_{\mathrm{zed}}(v)$ decays exponentially, the integrals in
Eqs.~(\ref{eq-betadef}) and~(\ref{eq-gammadef}) give finite values;
hence $\beta_{\mathrm{zed}}(L)=O(L^{-d})$ and
$\gamma_{\mathrm{zed}}(L)=O(L^{-3d})$.  If $n_{\mathrm{zed}}(v)$
decays as $v^{-x}$ with $x>2$, we have
$\beta_{\mathrm{zed}}(L)=O(L^{-d\min (1,x-2)})$ and
$\gamma_{\mathrm{zed}}(L)=O(L^{-d\min (3,x-2)})$.

On the other hand, if
the
order-parameter distribution $P(q)$ has a non-trivial part
as $L\to\infty$, $\alpha_{\mathrm{zed}}(L)$,
$\beta_{\mathrm{zed}}(L)$, and $\gamma_{\mathrm{zed}}(L)$ all give
finite values as $L\to\infty$.  These conditions are satisfied if the
distribution function decays as $n_{\mathrm{zed}}(v)\sim v^{-2}$.

\begin{figure}
\begin{minipage}[t]{0.49\textwidth}
\begin{center}
\includegraphics[width=8cm]{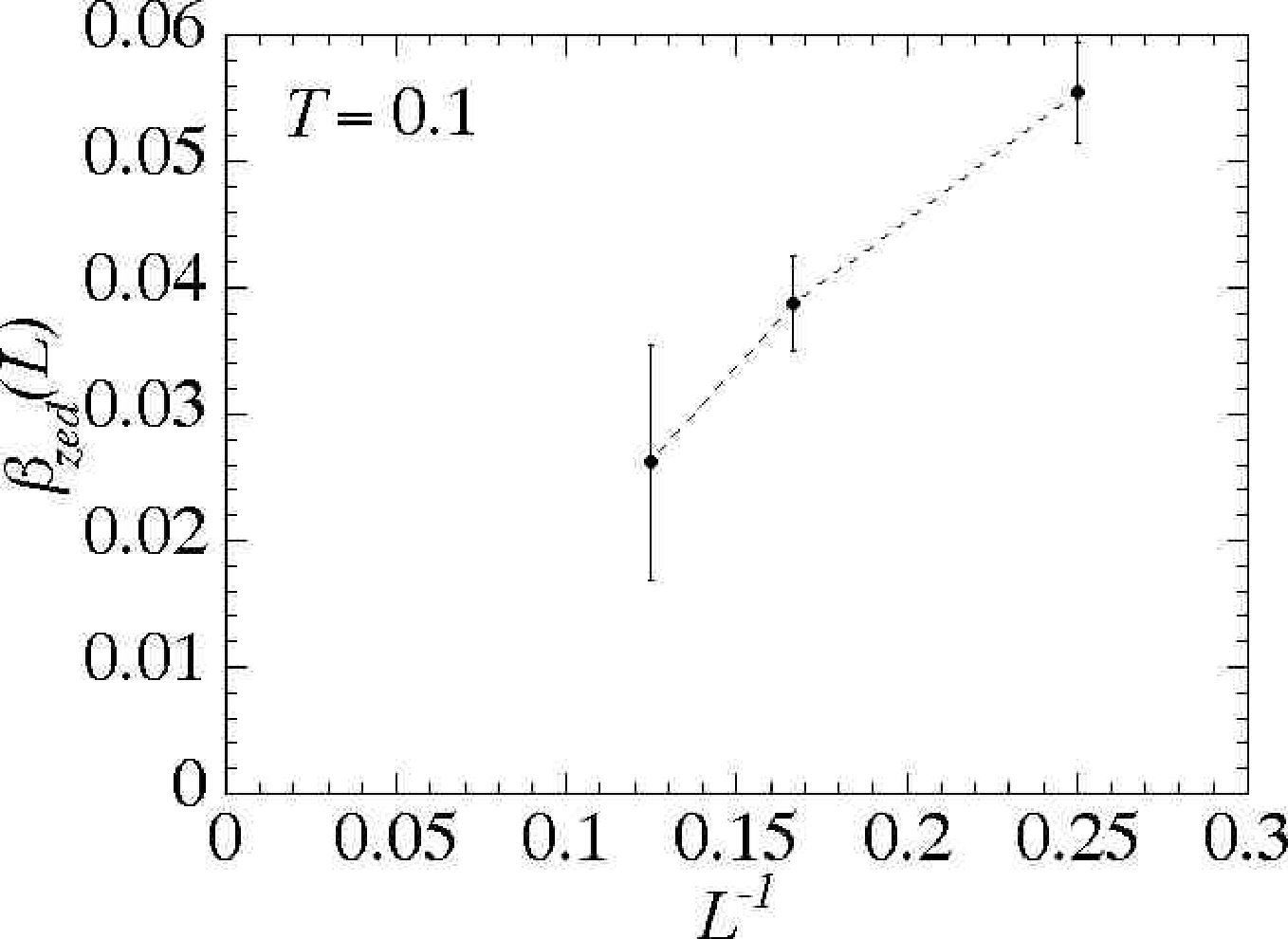}

(a)
\end{center}
\end{minipage}
\hfill
\begin{minipage}[t]{0.49\textwidth}
\begin{center}
\includegraphics[width=8cm]{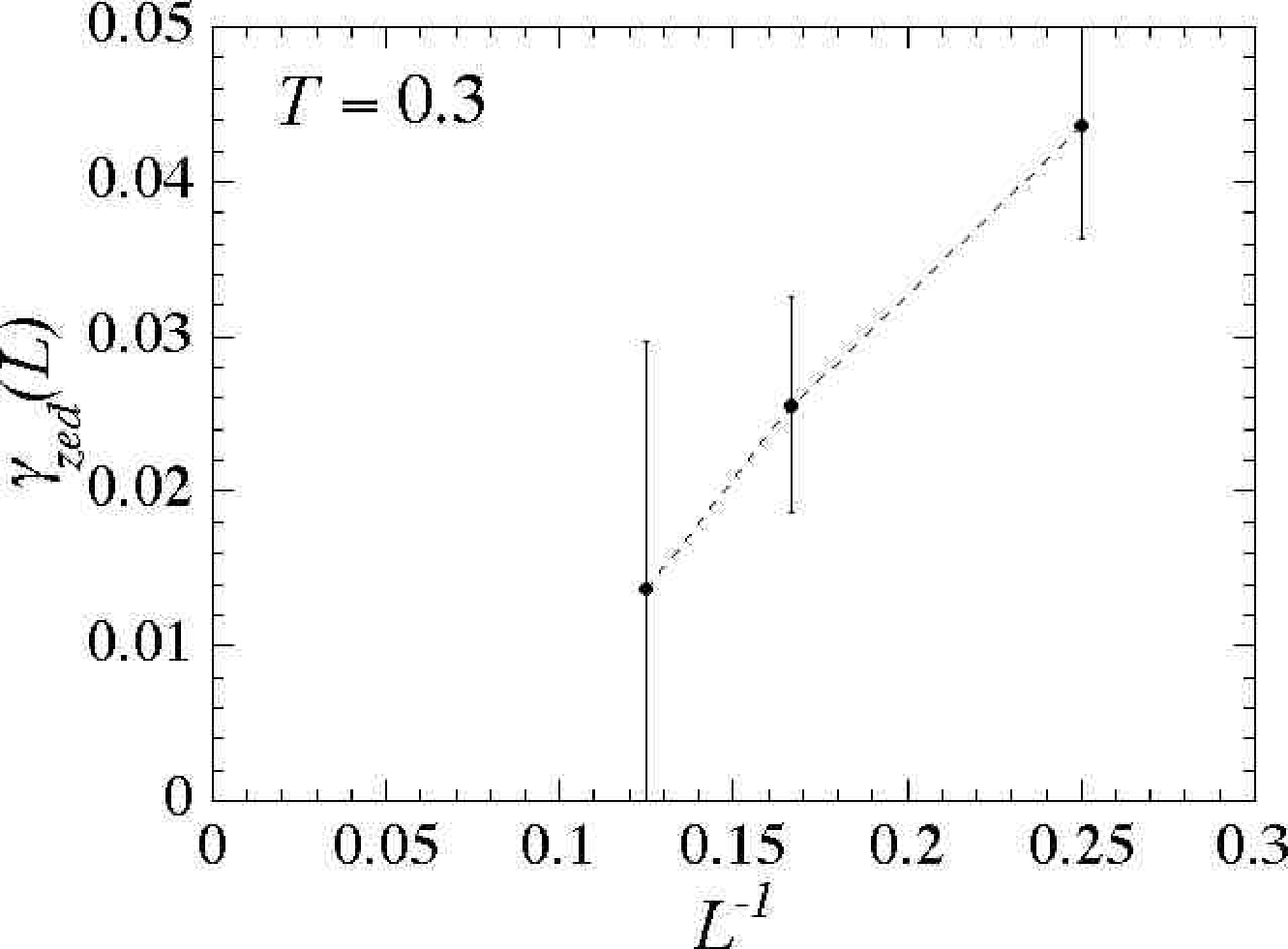}

(b)
\end{center}
\end{minipage}
\caption{The size dependence of (a) $\beta_{\mathrm{zed}}(L)$ and (b)
$\gamma_{\mathrm{zed}}(L)$. Both quantities are vanishing as $L\to\infty$.}
\label{fig-beta-gamma}
\end{figure}

In our numerical results, both $\beta_{\mathrm{zed}}(L)$ and
$\gamma_{\mathrm{zed}}(L)$ appear to be decreasing as $L$ is increased
(Fig.~\ref{fig-beta-gamma}).
The decrease is roughly $O(L^{-1})$, although the statistical errors
are too large to make a more definite statement.  This behavior is
additional evidence for a doubly degenerate ground state.  From our
numerical
results we suggest that the size distribution of the zero-energy
droplets decays as $n_{\mathrm{zed}}(v)\sim v^{-x}$ with $x>2$.

\section{Summary}\label{sec5}

The results of our multicanonical Monte Carlo calculation suggest a
doubly degenerate ground state.
Our main
findings supporting double degeneracy are: (i) $P(q)$ near $q\simeq0$
decreasing at low temperatures as the system size is increased; (ii)
the Binder parameter approaching unity at low temperatures; (iii)
the effect of the ground-state degeneracy on moments of the overlap
order parameter. In different temperature ranges, however, we further
demonstrated, within finite size limitations, that the same density of
states gave different scaling behaviors as a function of the system
size. Only at very low temperatures is the scaling consistent with a
doubly degenerate ground state.

We believe our numerical results are consistent with recent analytic
and numerical work and spin-glass themes pointing to a decoupling of
the degeneracy of the ground state from the standard droplet and
mean-field pictures and placing emphasis on the nature of the
low-energy excitations and entropic fluctuations with them. Our
numerical work provides information on these excitations only through
the density of states. The $P(0)$ predicted from this density of
states is trivial at low temperatures. For finite systems, our density
of states predicts crossover thermodynamic behavior as the temperature
is lowered.


\begin{acknowledgments}
We thank E. Domany for a helpful conversation. We gratefully
acknowledge computational support from the Applied
Mathematics program of the Department of Energy for the use of the
massively parallel computers at NERSC, from the Center for Nonlinear
Studies (CNLS) at Los Alamos National Laboratory for the use of its
parallel computer Avalon, and from the Los Alamos Directed Research
program at Los Alamos for computer time on the ASCI Blue Mountain
computer.  The use of the random-number generator library SPRNG of
National Center for Supercomputing Applications (NCSA) at University
of Illinois is also gratefully acknowledged.
A part of the computation was also carried out on a PC-cluster parallel
computer ARK of Aoyama Gakuin University~\cite{Nakata00}.
\end{acknowledgments}

\appendix

\section{Multicanonical Monte Carlo Method}\label{sec4}

We will  describe the multicanonical Monte Carlo method
from a viewpoint~\cite{Gubernatis00} slightly different from other
authors~\cite{Berg91,Berg92,Berg94,Berg96,Berg98}. We first
compare it with random-sampling and canonical methods.

Suppose that we perform a Monte Carlo simulation aimed at sampling from
the weight $W(E)$. After the simulation has equilibrated, that is,
after the Markov chain reaches a steady-state, each microscopic state
(spin configuration) is generated at a rate proportional to $W(E)$.
Also suppose that in ``equilibrium'' we construct a histogram $h(E)$ of
the energy $E$ values associated with each configuration (state)
generated.  The rate at which a state with the energy $E$ appears is
proportional to $D(E)W(E)$, where $D(E)$ is the density of states.

\begin{figure}
\begin{center}
\includegraphics[width=8cm]{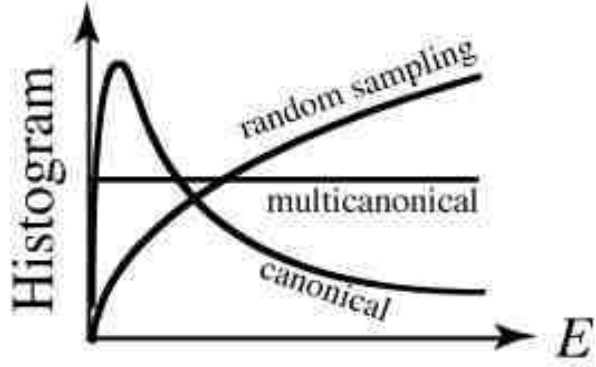}
\end{center}
\caption[]{Schematic of the histograms of energy values
generated in the random-sampling, canonical, and multicanonical
simulations.}
\label{fig-hist}
\end{figure}

In a random-sampling simulation, one generates all microscopic states
at the same rate.  In other words, $W(E)=\mathrm{const.}$, and hence
the resulting histogram is proportional to the density of states:
$h(E)\propto D(E)$ (see Fig.~\ref{fig-hist}).
In many cases, the density of states becomes very small in low-energy
regions.  Hence simulations based on the random sampling
generally have difficulty investigating low-temperature properties
because the low-energy states, which dominate the thermodynamic average
at low temperatures, are generated only infrequently, if at all.

Importance sampling algorithms, like the Metropolis algorithm for
canonical ensemble, were developed to overcome this difficulty.  Here
states are generated with the importance weight $W(E)\propto e^{-\beta
E}$, and hence the energy histogram is $h(E)\propto e^{-\beta E}D(E)$.
Thus, in ideal cases, low-energy states appear at a greater frequency
at low temperatures.  In spin-glass simulations, however, the
simulation tends to get stuck in local minima of the free-energy
landscape.

In the multicanonical simulations, the importance weight is set to
$W(E) \propto 1/D(E)$, and hence the energy histogram should be flat.
Of course, one cannot set $W(E)\propto 1/D(E)$ because the density of
states is {\it a priori} unknown.  However, the multicanonical method
is a procedure that makes the importance weight converge to the
reciprocal of the density of states~\cite{Smith95}.  The aim is to
ensure better statistics of low-energy states than in the
random-sampling method and, at the same time, generate more
high-energy states than in the canonical method so that the system can
escape local free-energy minima.

Berg and others have invented~\cite{Berg91,Berg92} and
applied~\cite{Berg94,Berg96,Berg98}
this method to spin glasses, taking the weight of each spin
configuration $\{\sigma\}$ either as
\begin{equation}\label{eq2-70}
W(\{\sigma\}) \propto {1}\bigm/{D\left(E(\{\sigma\})\right)}
\end{equation}
or
\begin{equation}\label{eq2-80}
W(\{\sigma\}) \propto {1}\bigm/{D\left(q(\{\sigma\})\right)},
\end{equation}
where $D(E)$ is the density of states with respect to the energy and
$D(q)$ is the one with respect to the overlap order parameter.

In the ideal situation, the multicanonical simulation effectively
generates a random walk in the macroscopic phase space. Because all
the thermodynamic states should appear with the same probability, one
can expect that the autocorrelation time of a thermodynamic variable
($E$ in the above case) is proportional to the volume (the number of
the sites).  In the simulation of the $\pm J$ model, the energy change
after each spin flip is of the order of $J$.  On the other hand, the
upper and lower bounds of the energy are of the order of $NJ$, where
$N=L^d$ is the number of the spins.  Hence the number of spin flips
necessary for the random walk to cover the entire energy phase space
is on the average of the order of $N^{2}$.  Thus if random walk is in
energy space, a one-dimensional space, the measured energy should
decorrelate after $N^{2}$ Monte Carlo steps or $N$ Monte Carlo sweeps.
(A sweep is the attempt to flip each Ising spin once on average.)

This ideal situation was not achieved in the previous
applications of the multicanonical simulations to spin
glasses~\cite{Berg94,Berg96,Berg98}.  Berg {\it et al}.\ reported
$\tau\propto N^{2.8}$ for the importance
weight~(\ref{eq2-70})~\cite{Berg94} and $\tau\propto N^{2.42}$ for the
importance weight~(\ref{eq2-80})~\cite{Berg98}.  (Here $\tau$ is
measured in the unit of Monte Carlo {\em sweep}.)  This suggests that
the slow dynamics in the low-energy region was not removed completely
in these mono-variate multicanonical simulations.  (Note, however, that
Berg {\it et al}.\ did not use the conventional definition of the
autocorrelation time in measurement of $\tau$.  This might be a
contribution to the above power law behavior.)

\begin{figure}
\begin{minipage}{0.45\textwidth}
\includegraphics[width=\textwidth]{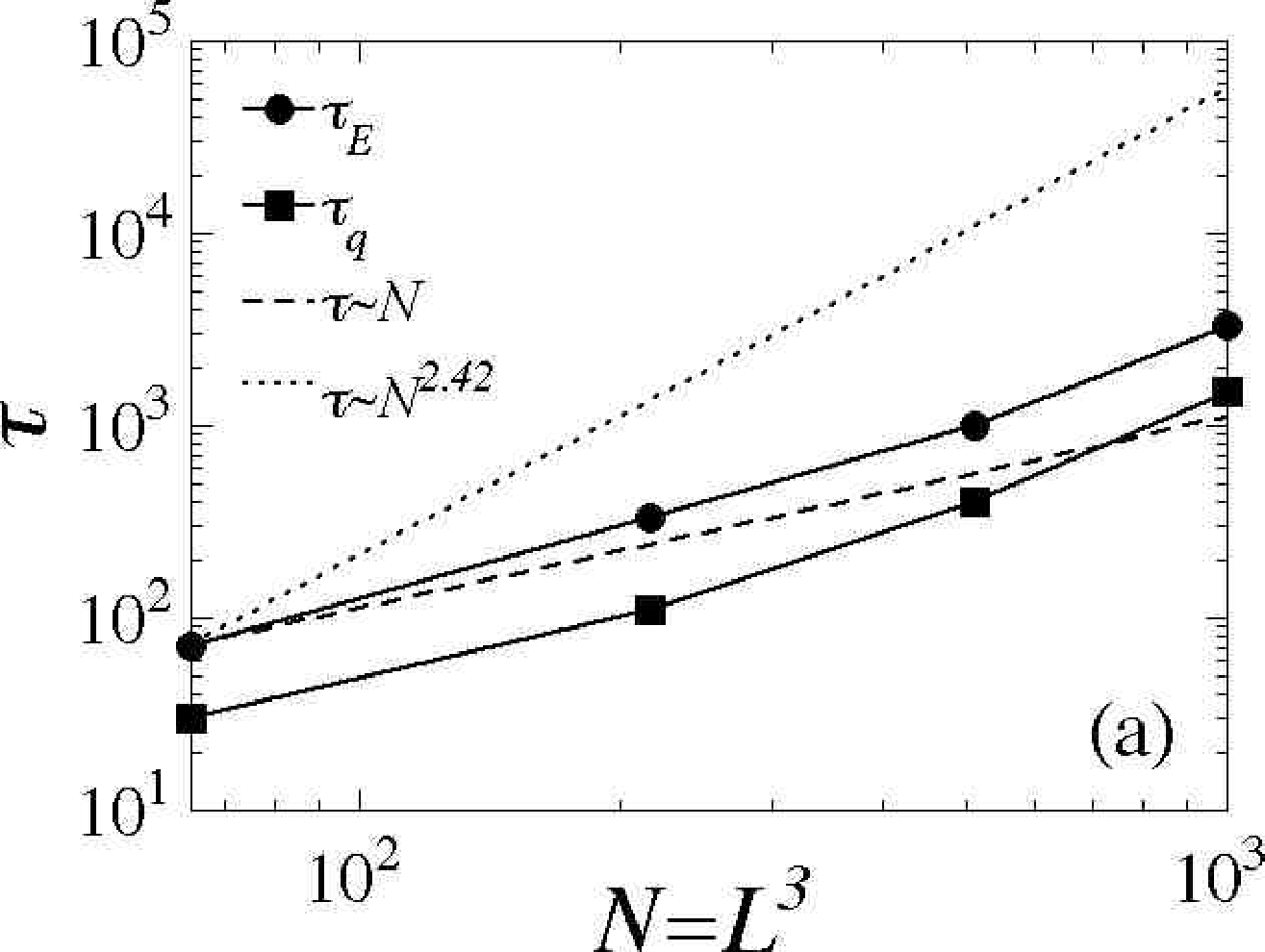}
\end{minipage}
\hfill
\begin{minipage}{0.45\textwidth}
\includegraphics[width=\textwidth]{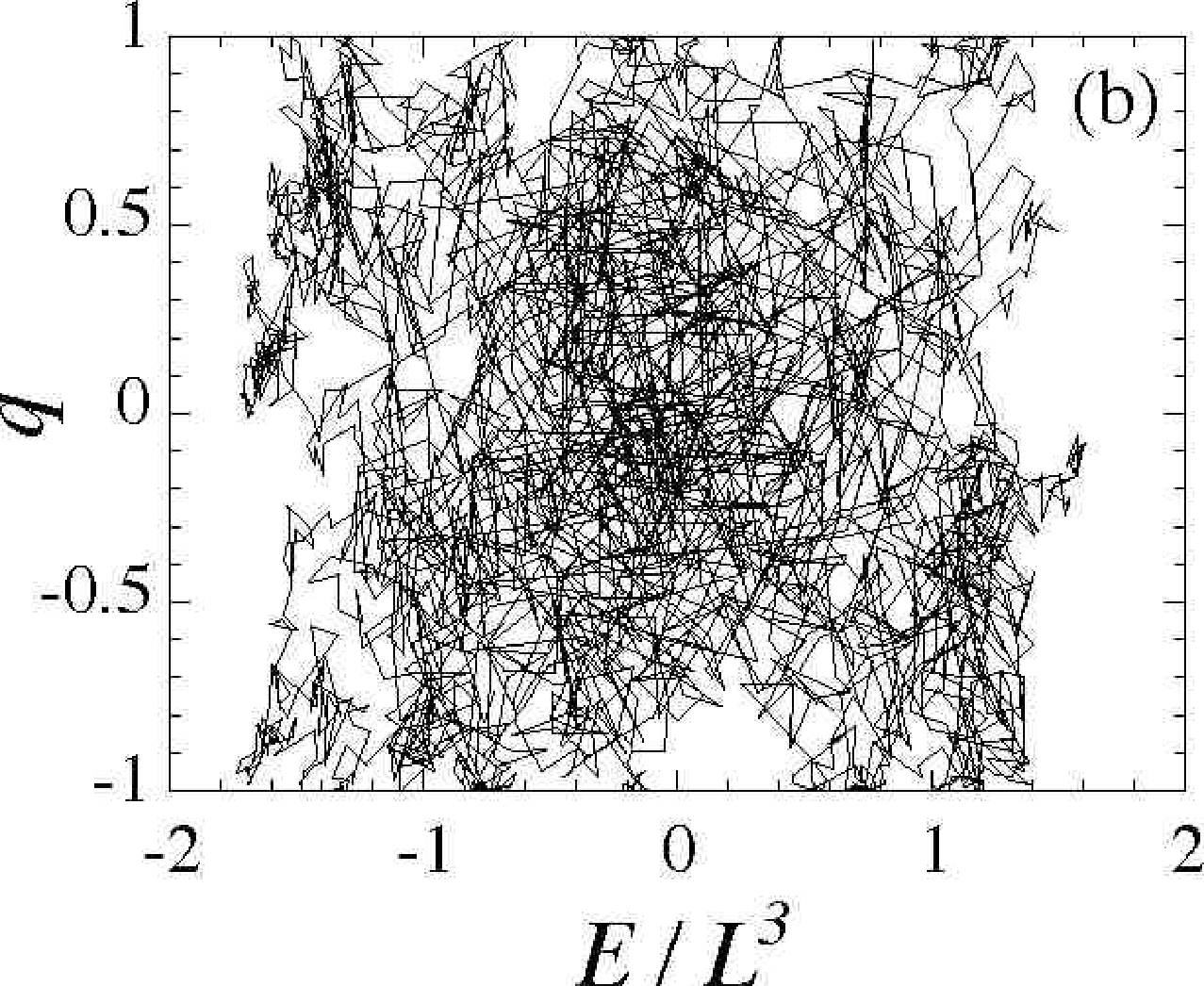}
\end{minipage}
\caption[]{(a) System-size dependence of the autocorrelation time
$\tau$ with respect to $E$ and $q$.  The dotted line represents the
power law $\tau\propto N^{2.42}$, which was reported
in~\protect\cite{Berg98}.  (b) An example of the random walk in the
phase space $(E,q)$. The system size simulated here is $L=4$. The time
series of 2500 Monte Carlo sweeps after equilibration is plotted.}
\label{fig-correl}
\end{figure}

In the present study, we carried out the simulation using the
bivariate importance weight:
\begin{equation}\label{eq2-90}
W\propto {1}\bigm/{D(E,q)}.
\end{equation}
As shown in Fig.~\ref{fig-correl} (a), we almost achieved the ideal
situation $\tau\propto N$.
The random walk in the two-dimensional phase space is illustrated in
Fig.~\ref{fig-correl} (b).
The autocorrelation plotted in Fig.~\ref{fig-correl} (a) was computed
as follows:
We first measured the autocorrelation as the simple Monte Carlo average
without any weighting:
\begin{eqnarray}
C_E(t)
&\equiv&
\frac{1}{N_{\rm MCS}}\sum_{j}E(j)E(j+t)
\\
C_q(t)
&\equiv&
\frac{1}{N_{\rm MCS}}\sum_{j}q(j)q(j+t),
\end{eqnarray}
where $N_{\rm MCS}$ is the number of Monte Carlo steps for
measurement.
Then the autocorrelation time was computed as the integrated time:
\begin{eqnarray}
\tau_E\equiv
\frac{1}{C_E(0)}\sum_{t=0}^{\tilde{t}_E}C_E(t),
\\
\tau_q\equiv
\frac{1}{C_q(0)}\sum_{t=0}^{\tilde{t}_q}C_q(t),
\end{eqnarray}
where $\tilde{t}_E$ ($\tilde{t}_q$) is defined as the time when
$C_E(t)$ ($C_q(t)$) first becomes negative.

There is a drawback with the bivariate simulation: large electronic
memory is needed to store the data from which the bivariate histogram
is constructed. For the Ising model we used the number of all possible
discrete states to equal the number of bins in the energy phase space.
This number was of the order of $L^{d}$ and was similarly sized in the
$q$ phase space.  Hence an array of the size $L^{2d}$ was necessary to
store each density of states $D(E,q)$, the histogram $h(E,q)$, and
related work space.  The combined storage requirements amounted to
nearly 30MB per sample for $L=10$ and 400MB per sample for
$L=16$. Based on direct comparisons with the original mono-variate
multicanonical method, we have, however, concluded that for spin-glass
simulations the use of both $E$ and $q$ is essential for improving the
slow dynamics~\cite{Hatano-SG1}.  We comment that bivariate
multicanonical Monte methods have been previously used in different
contexts, {\it e.g.\/}, protein folding
simulations~\cite{bivar-MCMC1,bivar-MCMC2}.

The actual algorithm of the multicanonical method has the following
structure:
\renewcommand{\labelenumi}{(\Alph{enumi})}
\renewcommand{\labelenumii}{(\alph{enumii})}
\renewcommand{\labelenumiii}{(\roman{enumiii})}
\begin{enumerate}
   \item\label{en1}
      Make a guess at the density of states, $D_{0}(E,q)$.
      Start the following loop with $i=0$.
   \item\label{en2}
    The outer loop (multicanonical-iteration loop):
   \begin{enumerate}
     \item\label{en2-1}
        Set the importance weight to $W_{i}=1/D_{i}$.
        Calculate the energy
        and the order parameter of the initial spin
        configuration, $E_{0}$ and $q_{0}$.
        Start the following loop with $j=0$.
     \item\label{en2-2}
        The inner loop (spin-update loop):
     \begin{enumerate}
        \item\label{en2-2-1}
           Choose a spin to be updated. Calculate the energy and the
           order parameter, $E_{j}'$ and $q_{j}'$,
           assuming that the spin is actually flipped.
        \item\label{en2-2-2}
           Calculate the spin-flip probability as
           \begin{equation}\label{eq3-100}
             P_{\mathrm{flip}}
 =\min\left(1,\frac{W_{i}(E_{j}',q_{j}')}{W_{i}(E_{j},q_{j})}\right)
           \end{equation}
        \item\label{en2-2-3}
           Draw a random number $R\in[0,1]$. If $R<P_{\mathrm{flip}}$, flip
the
           spin and make the substitution $E_{j+1}=E_{j}'$ and
$q_{j+1}=q_{j}'$.
           Otherwise, keep the current spin configuration and make
            the substitution $E_{j+1}=E_{j}$ and $q_{j+1}=q_{j}$.
        \item\label{en2-2-4}
           Increment the histogram
           bin $h_{i}(E_{j+1},q_{j+1})$ by one.
            This step should be skipped during the burn-in stage 
(specifically
            given below), that is, until the
            simulation reaches ``equilibrium'' for the importance weight 
$W_{i}$.
       \item\label{eq2-2-5}
           Increment the number of steps $j$ by one and go to the
           step~(i) until $j$ exceeds a pre-determined number.
      \end{enumerate}
      \item\label{en2-3}
         Measure physical quantities $Q(E,q)$ via
         \begin{equation}\label{eq3-110}
           \left\langle Q \right\rangle_{i}
           =\frac{1}{Z_{i}}\sum_{E,q} Q(E,q) e^{-\beta E} h_{i}(E,q)
           D_{i}(E,q),
         \end{equation}
         where the partition-function estimate is given by
         \begin{equation}\label{eq3-120}
           Z_{i}=\sum_{E,q} e^{-\beta E} h_{i}(E,q) D_{i}(E,q).
         \end{equation}
         In practice, this step was skipped until the histogram $h_{i}$ 
became
         reasonably flat (see below for more specifics).
       \item\label{eq2-4}
         Update the guess of the density of states as
         follows:~\cite{Smith95}
         \begin{equation}\label{eq3-130}
           D_{i+1}(E,q)=\frac{D_{i}(E,q)\left(h_{i}(E,q)+1\right)}%
           {\mathrm{Normalizaion}}.
         \end{equation}
         The normalization constant is chosen so that the integration
         of the density of states becomes unity.
         (An extra count is added to the histogram in~(\ref{eq3-130}) in
         order to avoid zero division in setting the next importance weight
         $W_{i+1}=1/D_{i+1}$~\cite{Smith95}.)
       \item\label{eq2-5}
         Increment the number of multicanonical iterations $i$ by
         one and go to the step~(a), until the histogram meets
         certain conditions of being flat.
    \end{enumerate}
\end{enumerate}

The method is an iterative procedure requiring an initial guess
$D_0(E,q)$ of $D(E,q)$. We found our average results to be quite
independent of this guess, but used the following particularly
convenient and efficient choice: Our choice was a flat one for
$L=4$, a case where the convergence was very rapid.
The results for this system size were then scaled and used for
the starting points for larger systems.

For a given iteration, step~(iv) should be skipped until the
simulation reaches ``equilibrium'' for the importance weight
$W_{i}$. For $L=4$, 6, 8, and 10, we used 10 000, 40 000, 50 000, and
100 000 Monte Carlo sweeps for this burn-in stage. The number of the
burn-in sweeps were approximately a hundred times longer than the
autocorrelation time given in Fig.~17a. The autocorrelation times were
determined before we began our production running.  After the burn-in
we used 1 000 000, 4 000 000, 5 000 000, and 10 000 000 sweeps for the
different systems sizes during each iteration step; that is, the
number of sweeps in each iteration was 10 000 times the measured
autocorrelation time and one hundred times the burn-in time.

Step~(c) was skipped until the histogram $h_{i}$ became reasonably
flat.  The number of iterations required for this varied as function
of $L$ and from one random bond configuration to another for a given
$L$. For $L=4$, the range was 2 to 15; for $L=6$, 2 to 23; for $L=8$,
2 to 276; and for $L=10$, 4 to 227. True flatness is never achieved
nor is it necessary. What is important is the histogram be reasonably
flat, extends to the band edges, and lacks spikes at the band
edges. In defining a ``reasonable'' flatness criteria, which is the
criterion for the iteration to have converged, we were motivated by
the following observations: While converging, the histogram typically
is first concentrated in the mid-part of the density of states and
expands outward towards the upper and lower bounds of the energy (band
edges) and the order parameter until it becomes flat.  At the leading
edges, spiky structures sometimes appear as the histogram is crossing
over from occupied to unoccupied states. When the edges are reached,
the spikes disappear. We defined the histogram to be flat, whenever
the total count at the minimum energy and that at $E=0$ were within
one sigma of the average histogram
count. In contrast to the usual Ising model, the band width for a
bond configuration is not known {\it a priori}, but it is easy to
estimate from short runs and then set it slightly larger than expected
to handle extremal states. In general, we found the multicanonical
method to be an efficient estimator of the lowest energy of a given
bond configuration.

After convergence, we repeated the outer loop five times after so that
we could detect erroneous estimates of the density of states; that is,
using the just determine $D(E,q)$ as the starting point, we repeated
multicanonical procedure until five consecutive iterations maintained
convergence. Testing determined that five was an adequate number to
avoid false convergence. We also divided the inner loop into ten
blocks, from which we calculated the statistical errors.
Because the length of each block is about 1000 times longer
than the auto-correlation time, the absence of correlations between
block averages is expected and standard estimates for statistical
error were used.

Because it is an importance estimator, (\ref{eq3-110}) could in
principle be applicable before the histogram becomes flat; that is,
{\em achieving flatness is not an essential requirement for the
accuracy of the method if importance weighted estimators are used.\/}
Even with convergence, we used such estimators for the sake of
caution.  {\em We only collected data for converged results\/.} With
sufficient statistics, the histogram $h_{i}$ would be proportional to
$W_{i}(E,q)D(E,q)=D(E,q)/D_{i}(E,q)$, where $D(E,q)$ is the true
density of states.  Thus the estimator~(\ref{eq3-110}) would give
\begin{equation}\label{eq3-200}
    \left\langle Q \right\rangle_{i}
      \simeq\frac{1}{Z_{i}}\sum_{E,q} Q(E,q) e^{-\beta E} D(E,q),
\end{equation}
with the partition function
\begin{equation}\label{eq3-210}
    Z=\sum_{E,q} e^{-\beta E} D(E,q).
\end{equation}

We comment again that we could have calculated all physical quantities after
the simulations were completed, provided we had stored the density of
states for each random bond configuration on disk.  This procedure
would requires a huge amount of disk space because of the need to
simulate a large number of random samples.

We also note that Marinari {\it et al}. \cite{Marinari_comment} have
recently tried to implement our bivariate version of Berg and
Neuhaus's multicanonical method and reported experiences very
different from ours. They reported difficulties in ergodicly sampling
the importance function, finding it only remained ``flat'' only by
using an order of magnitude larger number of Monte Carlo sweeps than
we used. This would correspond needing burn-in times and block sizes
1000 times our estimated auto-correlation time. More importantly the
very small values of $h_i$ erratically riddling their histograms would
translate into very small values of $h_i(E,q)D_i(E,q)$, which should
have caused our measured values for a given bond configuration to
vary widely. We did not observe this, and our estimates of
autocorrelation time were consistent for different system sizes. As
mentioned above, most of our data analysis was done ``on-the-fly'' to
avoid storage problems so we cannot re-examine the actual data for
our reported results. However, when a referee pointed out to us
the Marinari {\it et al}. difficulties, we immediately executed a
number of simulations for the $L=8$ system size causing Marinari {\it et 
al.}
problems and looked at the same quantity that they reported in their
Fig.~3.  We never saw the behavior that they claimed occurs. The
origin of the discrepancy between their results and ours is unknown to
us.


\begin{thebibliography}{99}
\bibitem{note2} Our preliminary results \cite{Hatano-SG1} showed a
different behavior because of a different and non-standard definition
of the Binder cumulant was used.
\end{thebibliography}

%
%

%
%

\end{document}